\documentclass[a4paper,11pt]{article}
\pdfoutput=1

\usepackage{jinstpub} 
\usepackage{subcaption}
\usepackage{gensymb}

\title{\boldmath Effect of variations in the gas mixture compositions on the timing and charge of glass RPC}

\author[a]{A. K. Sikdar,}
\author[a,b,1]{J. Sadiq, \note{Corresponding author.}}
\author[a]{and P. K. Behera}

\affiliation[a]{Physics Department, Indian Institute of Technology Madras, Chennai, India}
\affiliation[b]{Physics Department, Islamic University of Science and Technology, Kashmir, India}

\emailAdd{sadiqvengad@gmail.com}

\abstract
{
The India-based Neutrino Observatory (INO) is a  project aimed at building a large underground laboratory to explore the Earth's mater effects on the atmospheric neutrinos in multi-GeV range. INO will host a 50 kton magnetized iron calorimeter detector (ICAL) in which Resistive Plate Chambers(RPCs) will be the active detector elements. In ICAL, 28,800 glass RPCs of 2 m $\times$ 2 m size will be operated in the avalanche mode. A small variation in the compositions of ionizing gaseous medium in the RPC affects its performance. Study of the charge distribution of the RPC at different gas compositions is necessary to optimize the gas mixture. 

An RPC made with glass plates of dimension 30 cm $\times$ 30 cm was operated in avalanche mode with a gas mixture of $C_2H_2F_4$/$iC_4H_{10}$/$SF_6$. We have studied the performance of these RPCs at the same ambient conditions. The percentages of the $iC_4H_{10}$ or $SF_6$ were varied and its effect on the performance of RPC were studied. The study of the charge distribution and time resolution of the RPC signals at different gas compositions is presented in this paper. 
}


\begin{document}
\maketitle
\flushbottom

\section{Introduction} 
\label{sec:intro}

Resistive Plate Chambers (RPCs) have found its applications in the particle physics experiments as well as in medical fields \cite{santonico_2, rpc_medical}. Most of the high energy experiments prefer the RPCs as tracking detectors because of their excellent position and time resolutions \cite{santonico_rpcdev}. The India-based Neutrino Observatory (INO) has planned to build an underground laboratory which will house a 50 kton magnetised Iron CALorimeter (ICAL) detector \cite{icalwhitepaper}. The ICAL detector aims to determine the neutrino mass hierarchy and to improve the precision bounds on $\theta_{23}$ and $\Delta m^2_{32}$ from the Earth's mater effects on the atmospheric $\nu_\mu$ and $\bar{\nu}_\mu$ propagation. The ICAL will consist of 151 horizontal layers of 5.6 cm thick iron plates interleaved by 4 cm air gaps in which the detector elements will be placed. The Resistive Plate Chambers (RPCs) of dimension 2 m $\times$ 2 m are chosen as the active detector elements. Three identical modules, each of size 16 m $\times$ 16 m $\times$ 14.5 m, will be constructed.

RPCs consist of two parallel plate electrodes of high bulk resistivity ($\sim10^{12}~\Omega{cm}$) which provide a constant and uniform electric field in the gap between them when a high voltage ($\sim$ 10 kV) is applied across them. The gap contains a gaseous ionizing medium which is a mixture of R134a ($C_2H_2F_4$, 95$\%$), Isobutane ($iC_4H_{10}$, 4.5$\%$) and Sulphur hexafluoride ($SF_6$, 0.5$\%$). Freon acts as a secondary electron quencher with high probability for primary ionization and Isobutane acts as an absorber of UV photons. SF$_6$ is required to control the excess number of electrons in avalanche mode \cite{Kalmani_gassystem}. An incoming charged particle ionizes the gas medium followed by the propagation and multiplication of the $e^-$-ion pairs, which in turn induces a current signal on external pickup strips \cite{RPC_simulation_1}. The variations in the gas mixture compositions significantly affects the signal and hence the performance of RPCs. The effect of variation of $iC_4H_{10}$ or $SF_6$ on the charge distribution and time resolution of the RPC were studied.

\section{Experimental setup}

Two glass plates of thickness 3 mm manufactured by Saint-Gobain were cut in 30 cm $\times$ 30 cm dimensions, pasted with a conductive tape (T-9149) of uniform surface resistance on the outer surfaces, and then used to fabricate an RPC of 2 mm gap. The RPC was operated at 10.4 kV in avalanche mode with a mixture of three gases, viz. R134a ($C_2H_2F_4$), Isobutane ($iC_4H_{10}$) and Sulphur hexafluoride ($SF_6$). The RPC was sandwiched between two pick-up panels comprising of 10 copper strips each of width 28 mm, separated by a gap of 2 mm. The middle strip was read out to study the performance of RPC. 

\begin{figure}[htbp]
      \hspace{-5mm}
      \centering
      \includegraphics[width=0.95\linewidth,trim = 15mm 20mm 0mm 85mm, clip]{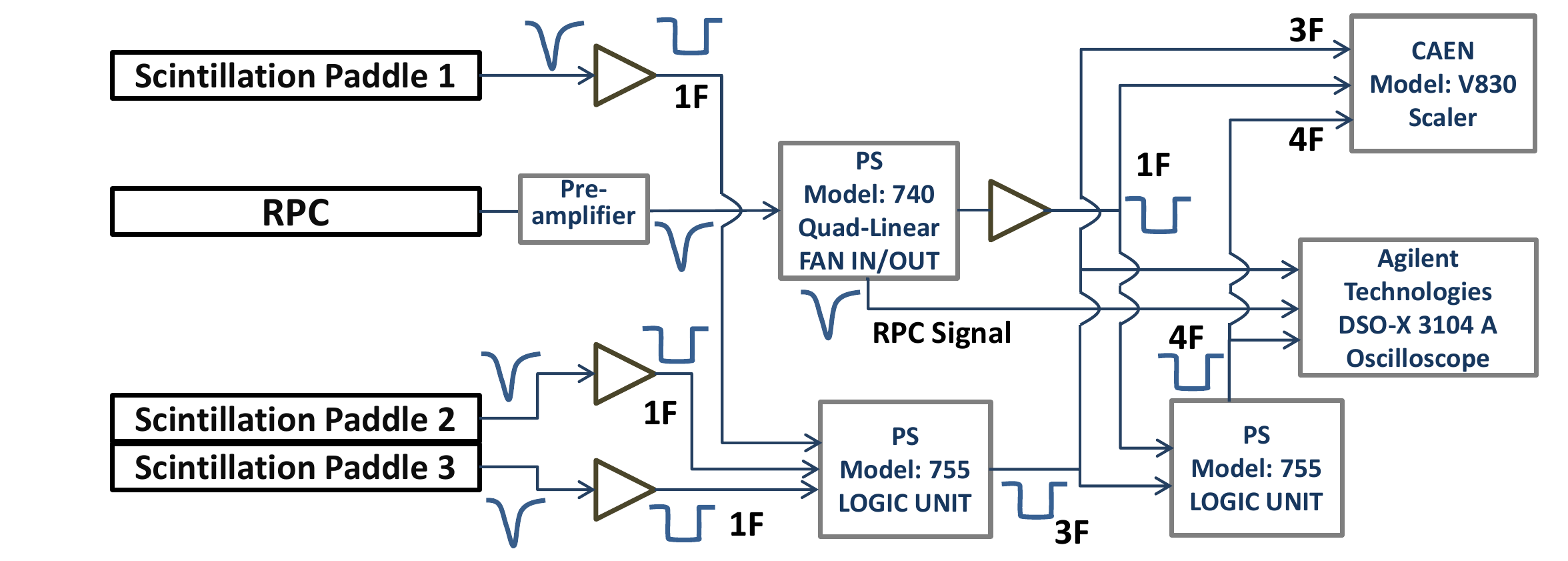}
      \vspace{-2mm}
      \caption{Schematic of experimental setup.}
      \vspace{-5mm}
      \label{fig:exp_setup}
    \end{figure}

Figure \ref{fig:exp_setup} shows the schematic of experimental setup to study the performance of the RPC. A cosmic  muon telescope was made using three plastic scintillator paddles P$_1$, P$_2$ and P$_3$ that are arranged vertically one above the other to get a 3-fold coincidence. The dimensions of P$_1$, P$_2$ and P$_3$ in length $\times$ width $\times$ thickness are 30~$\times$~2~$\times$~1~cm$^3$, 30 $\times$ 3 $\times$ 1 cm$^3$ and 30 $\times$ 5 $\times$ 1 cm$^3$ respectively. The RPC was placed in such a way that the central strip was aligned with the telescope window defined by the  paddles. A high voltage was applied across the electrodes of the RPC. The analog signals from the RPC were amplified with a preamplifier, since their amplitude is small in the avalanche mode of operation. The signals from the scintillator paddles and RPC were fed to the Data Acquisition System (DAQ). Scintillator paddle signals were converted to logic signals by a discriminator and ANDed in a logic unit to get the trigger pulse. The amplified RPC analog signal was connected to a linear FAN IN/FAN OUT (FIFO) to get two buffered output signals. One output signal was connected to an oscilloscope to measure the charge and time of the signal, and the other was converted to a logic signal by feeding it to a discriminator with a threshold voltage of -20 mV.

\subsection{Calibration of the gas system}

\begin{figure}[ht]
    \begin{subfigure}[b]{0.33\textwidth}
        \hspace{2mm}
        \includegraphics[width=1.10\textwidth,trim = 20mm 155mm 20mm 30mm, clip]{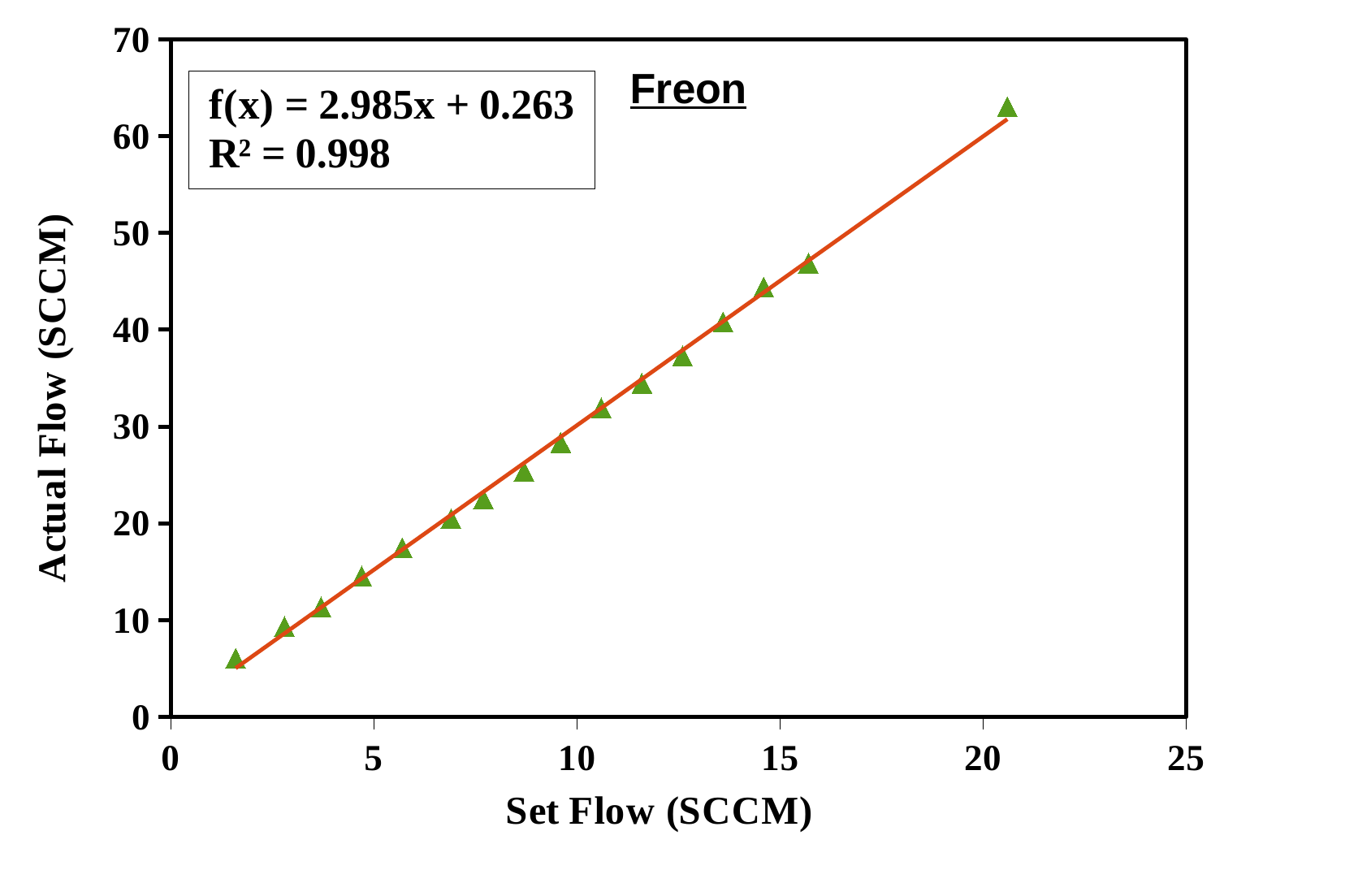}
    \end{subfigure}
    \begin{subfigure}[b]{0.33\textwidth}
        \hspace{1mm}
        \includegraphics[width=1.10\textwidth,trim = 20mm 155mm 20mm 30mm, clip]{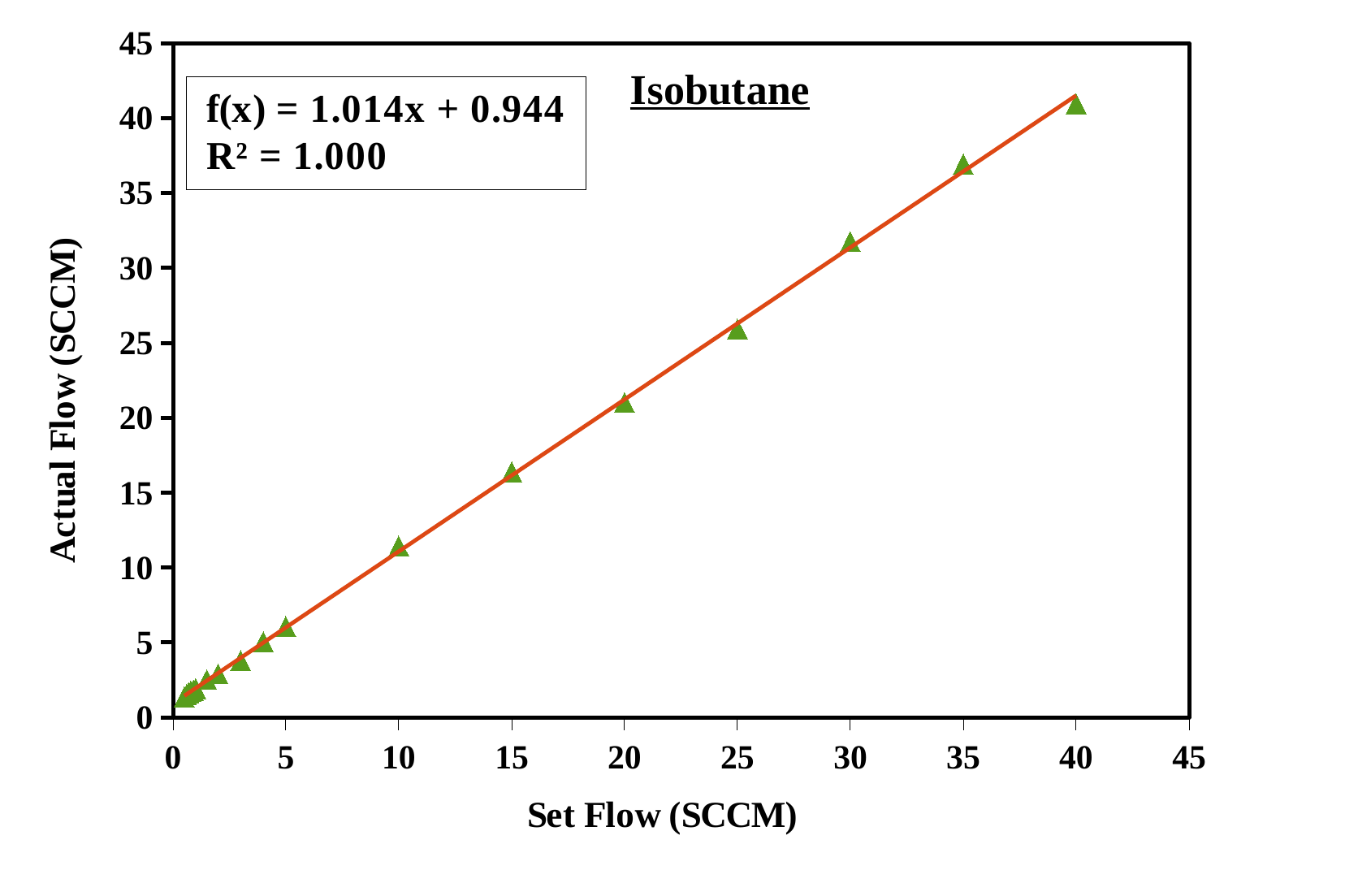}
    \end{subfigure}
    \begin{subfigure}[b]{0.33\textwidth}
        \hspace{2mm}
        \includegraphics[width=1.10\textwidth,trim = 20mm 155mm 20mm 30mm, clip]{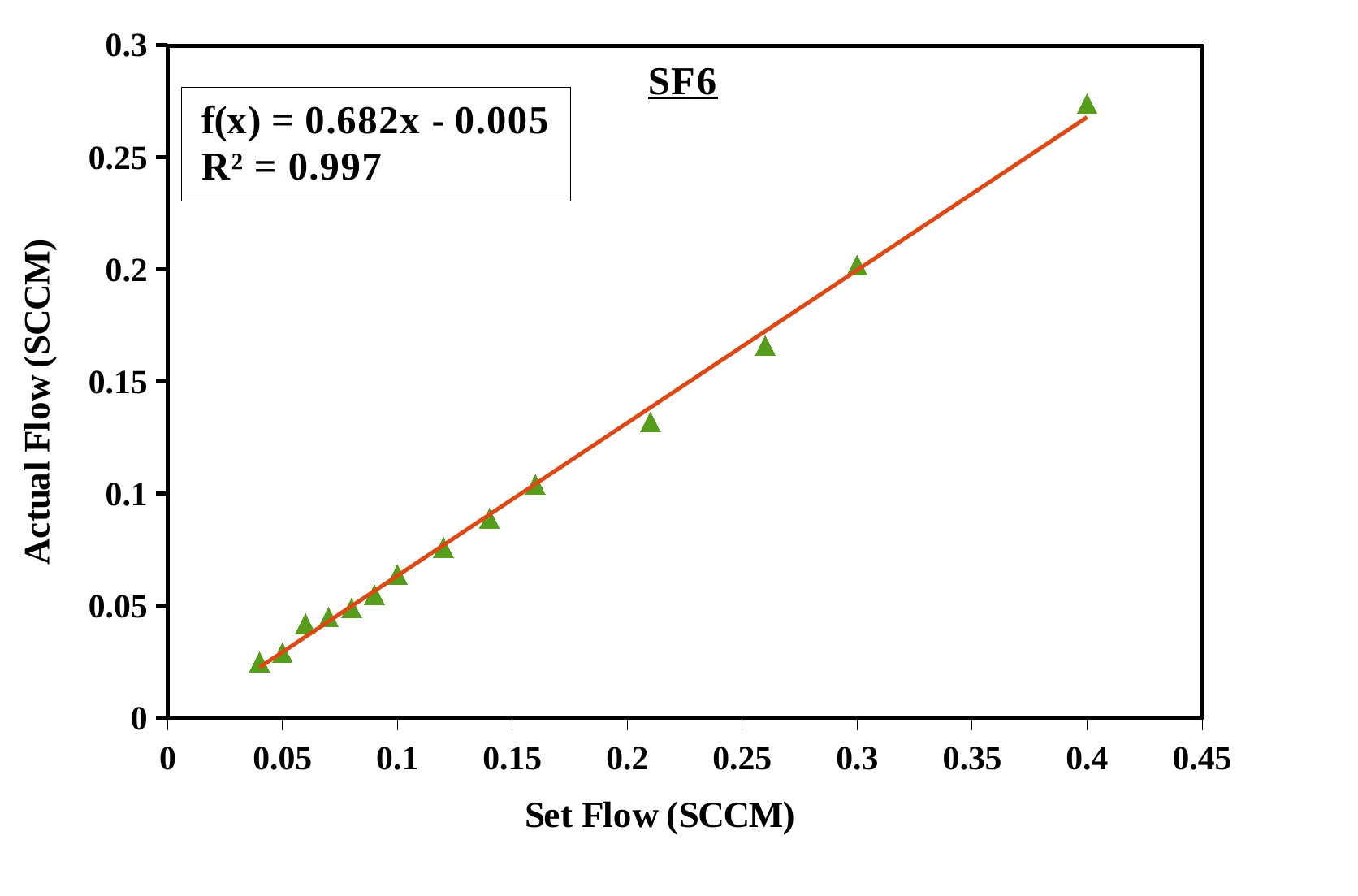}
    \end{subfigure}
    \vspace{-7mm}
    \caption{Calibration plots of MFCs corresponding to Freon, Isoutane and SF$_6$.}
    \label{gas_calibration}
    \vspace{-8mm}
   \end{figure}

Mixing of the gases in different proportions and their supply to the RPC are executed by an on-line gas mixing and multi-channel distribution system developed by INO team \cite{Kalmani_gassystem}. Mass Flow Controllers (MFCs) are used to control the flow of each gas at its required rate. Therefore, the MFCs were calibrated according to the required flow of the gases in the RPCs. All the MFCs were calibrated by measuring the time taken to reduce the water level in an inverted burette (water downward displacement method). The calibration plots of MFCs corresponding to Freon, Isoutane and SF$_6$ are shown in figure \ref{gas_calibration}. The required proportion of three individual gases in the mixture were determined, and the flow rates were adjusted in the corresponding MFCs according to the calibration. 

\section{Performance of RPC at different percentages of SF$_6$} 
SF$_6$ is a strong electronegative gas that captures free electrons. Therefore, addition of small amounts of SF$_6$ ($\sim$ 0.3 $\%$) significantly reduces the average pulse charge and suppresses the production of streamer signal in the RPCs \cite{Camarri1998}. To study the effect of variation of SF$_6$ concentration, we have flown the SF$_6$ at different proportions (from 0.3 $\%$ to 0.8 $\%$ in steps of 0.1 $\%$) fixing the proportion of $iC_4H_{10}$ at 15 $\%$. The charge induced on the copper strip was measured using an oscilloscope (DSO-X 3104 A) which stored the digitized analog pulses. Around 1000  signals were collected for each proportion of SF$_6$, and their charge distributions and time resolutions were studied.

\subsection{Charge distribution}

    \begin{figure}[ht]
        \vspace{-3mm}
        \hspace{-6mm}
        \centering
        \begin{subfigure}[b]{0.33\textwidth}
          \includegraphics[width=1.05\textwidth,trim = 0mm 0mm 0mm 0mm, clip]{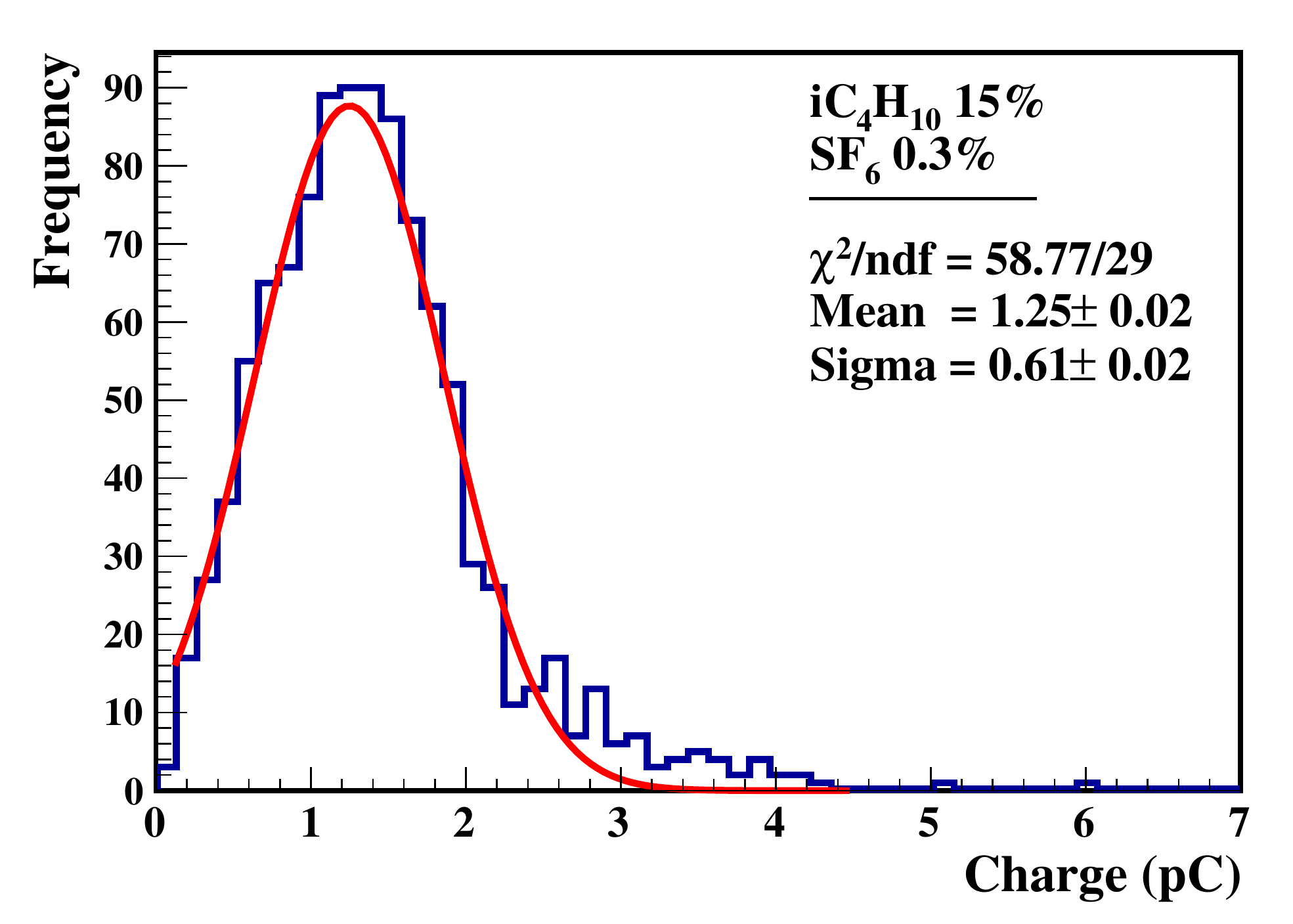}
        \end{subfigure}
        \begin{subfigure}[b]{0.33\textwidth}
          \includegraphics[width=1.05\textwidth,trim = 0mm 0mm 0mm 0mm, clip]{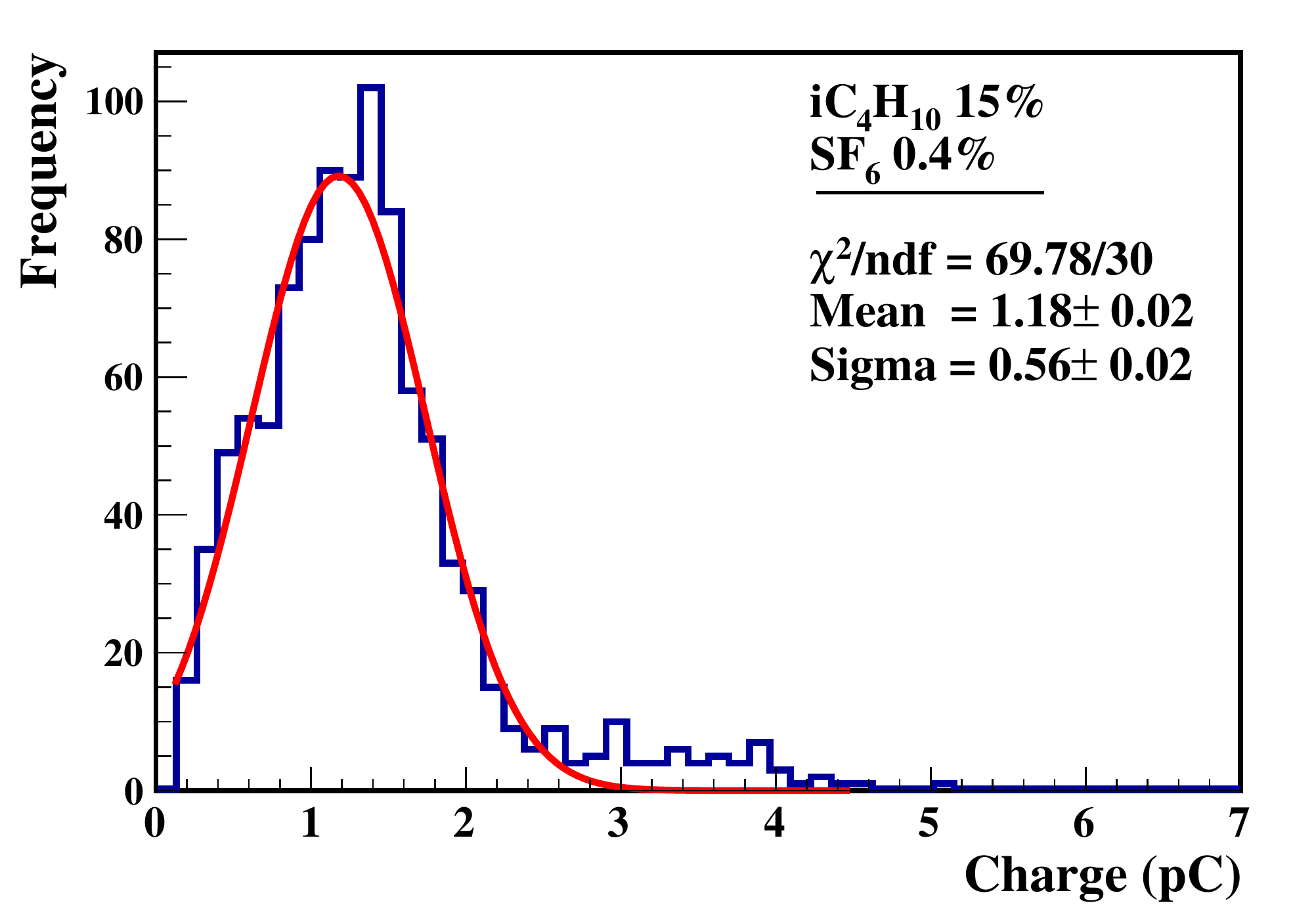}
        \end{subfigure}
        \begin{subfigure}[b]{0.33\textwidth}
          \includegraphics[width=1.05\textwidth,trim = 0mm 0mm 0mm 0mm, clip]{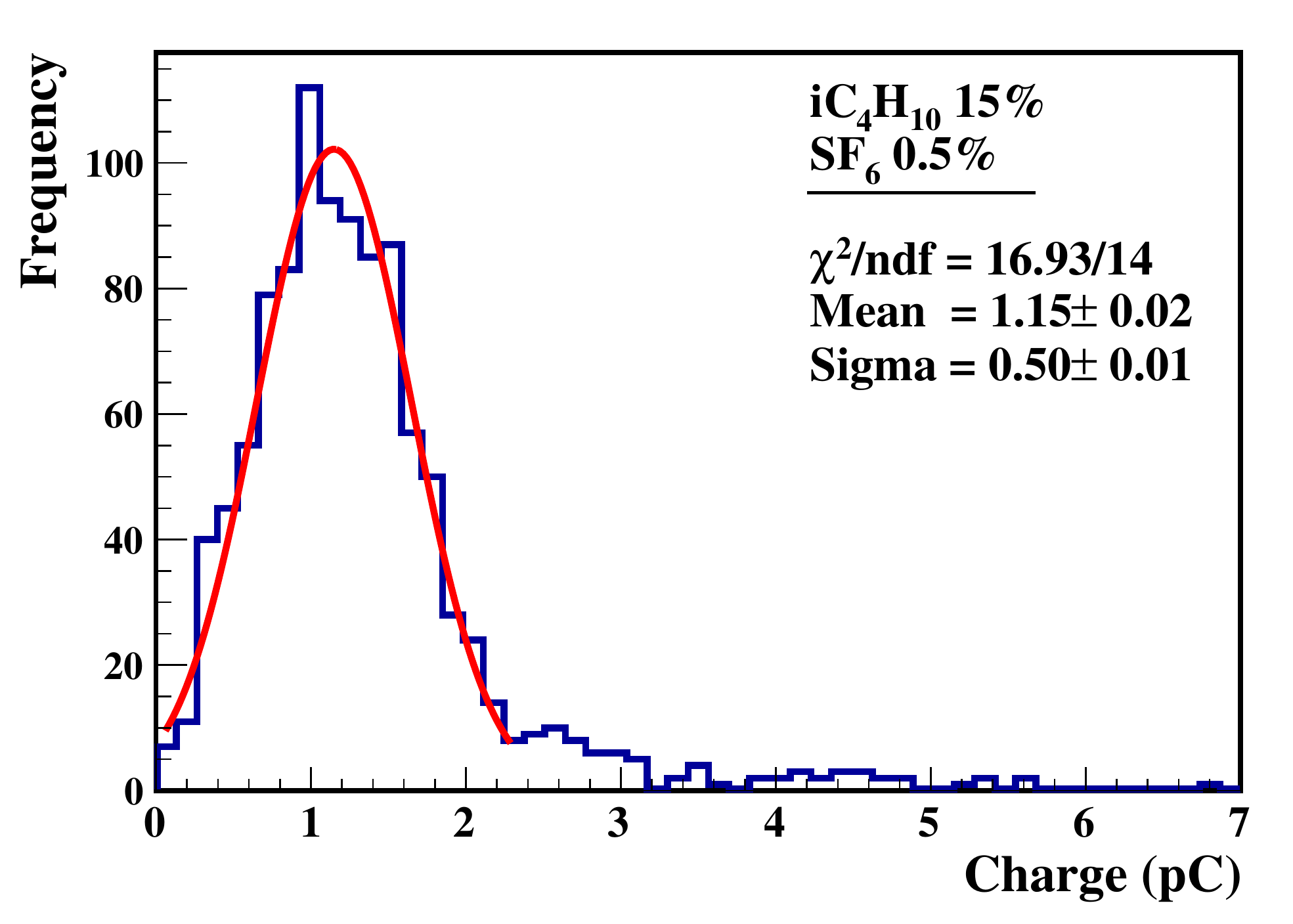}
        \end{subfigure}
        \hspace*{-6mm}
        \begin{subfigure}[b]{0.33\textwidth}
          \includegraphics[width=1.03\textwidth,height=0.161\textheight]{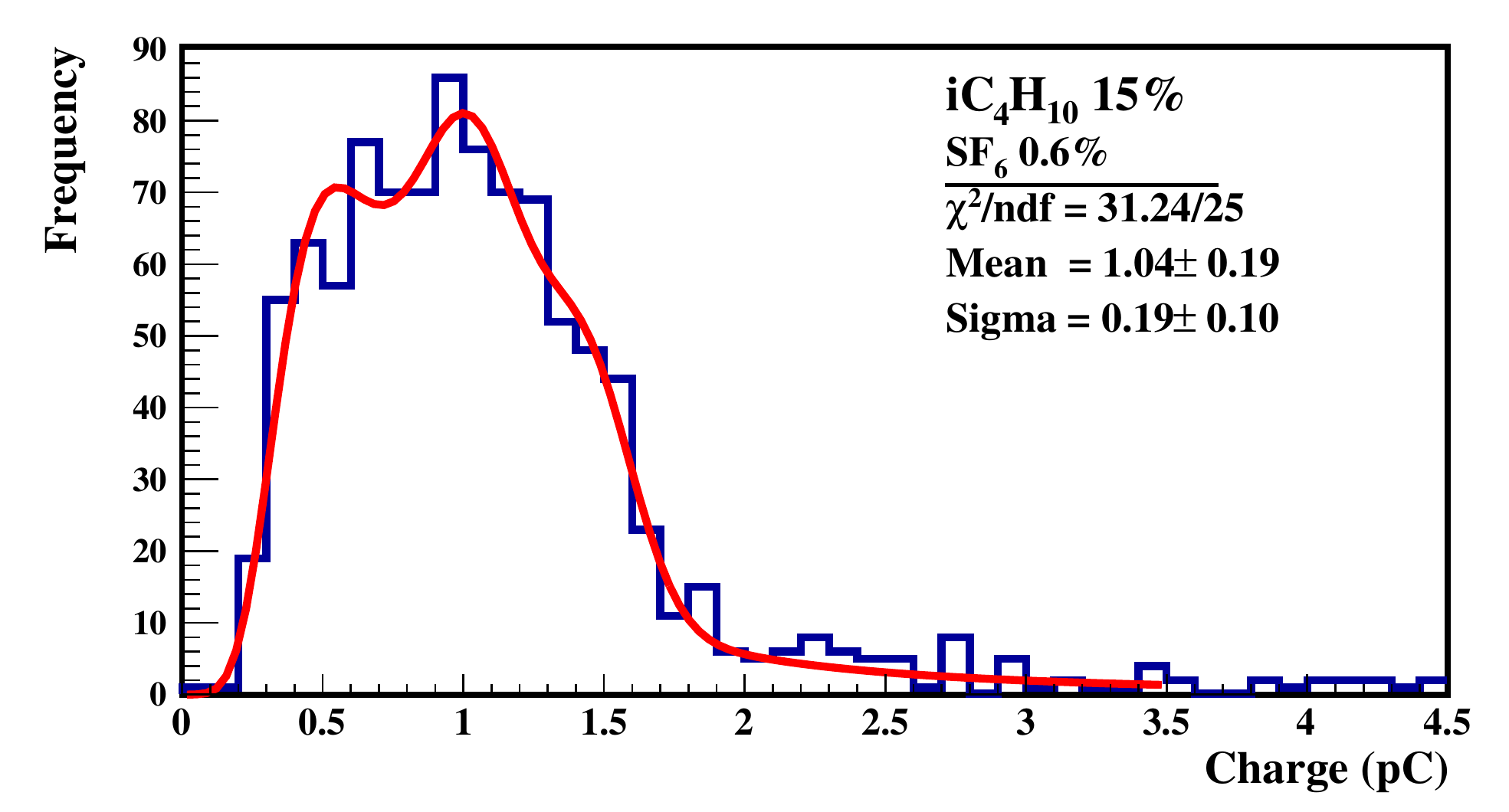}
        \end{subfigure}
        \begin{subfigure}[b]{0.33\textwidth}
          \includegraphics[width=1.03\textwidth,height=0.161\textheight]{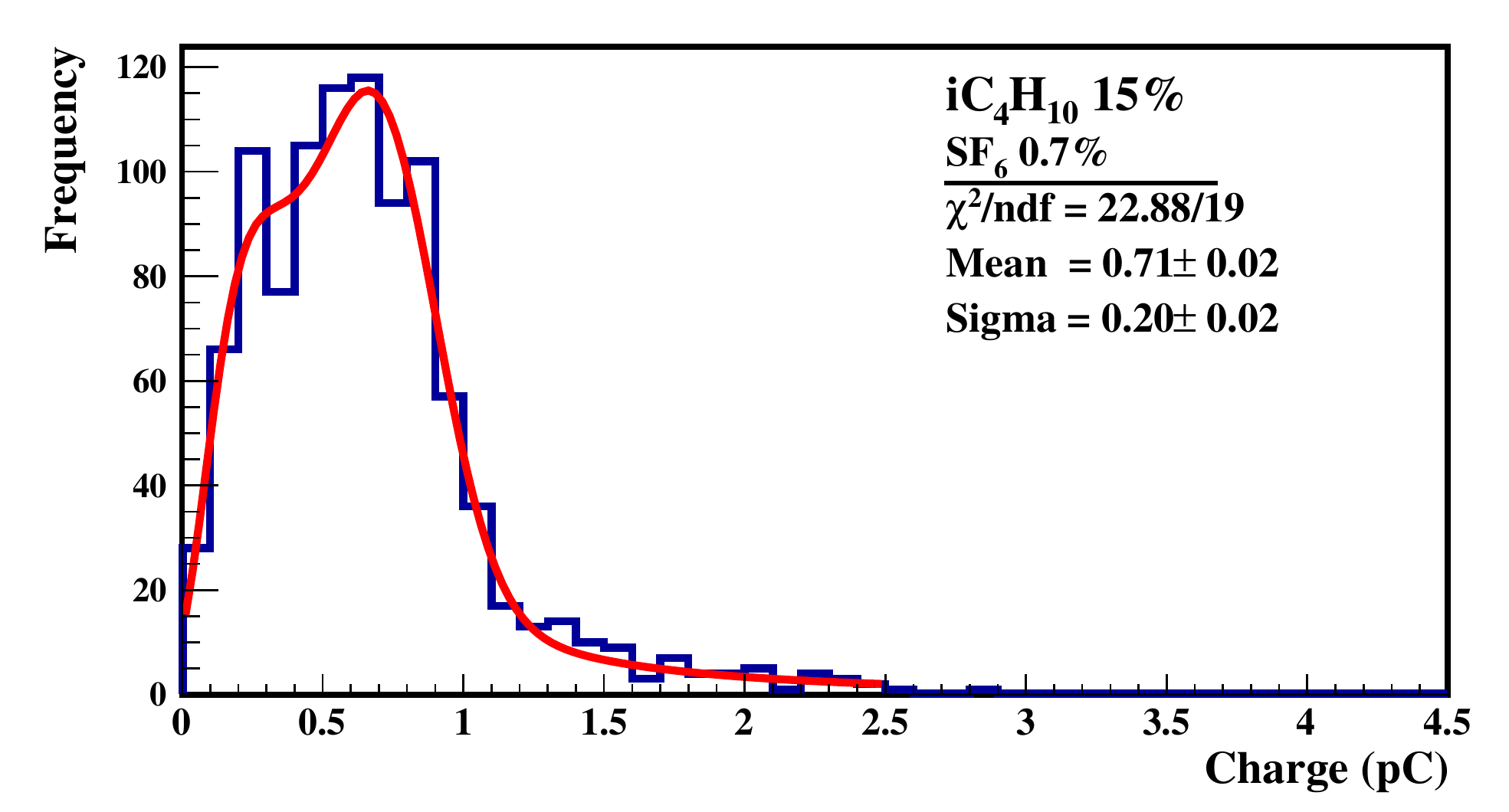}
        \end{subfigure}
        \begin{subfigure}[b]{0.33\textwidth}
          \includegraphics[width=1.03\textwidth,height=0.161\textheight]{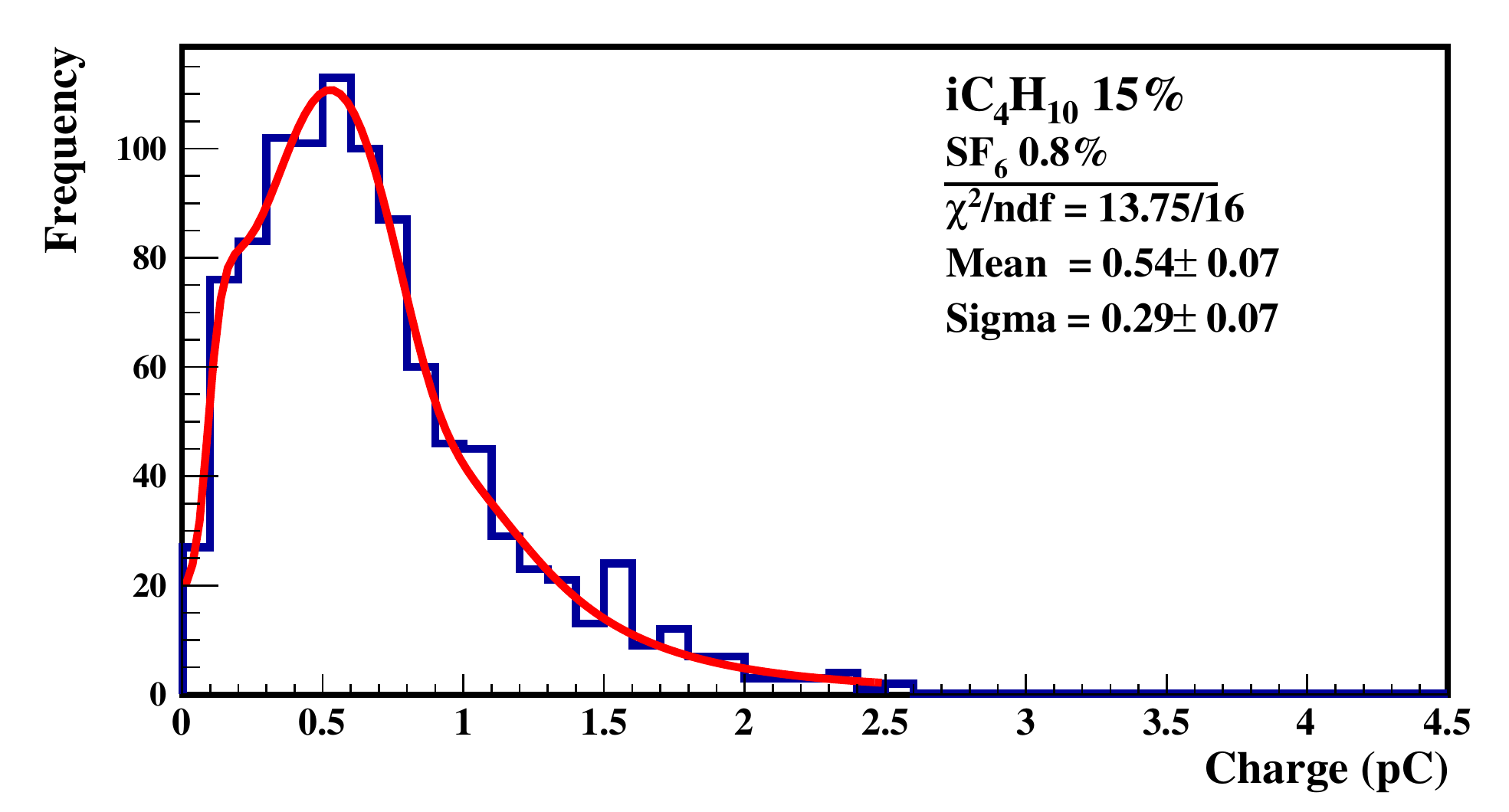}
        \end{subfigure}
        \vspace{-3mm}
        \caption{Charge distribution of the RPCs for various SF$_6$ concentrations at same operating voltage.}
        \label{fig:charge}
    \end{figure}

\begin{figure}[htbp]
        \hspace{1mm}
        \centering
        \begin{subfigure}[b]{0.49\textwidth}
          \includegraphics[width=1.0\textwidth]{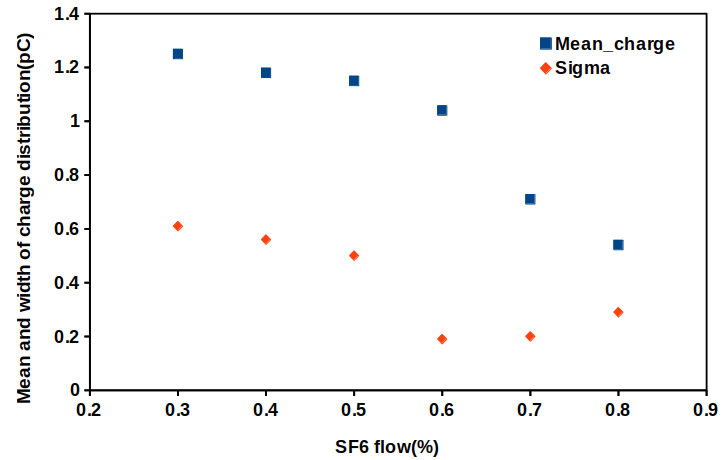}
          \caption{}
          \label{mean_charge_vs_SF6}
        \end{subfigure}
        \begin{subfigure}[b]{0.49\textwidth}
          \includegraphics[width=1.05\textwidth,trim = 20mm 162mm 20mm 30mm, clip]{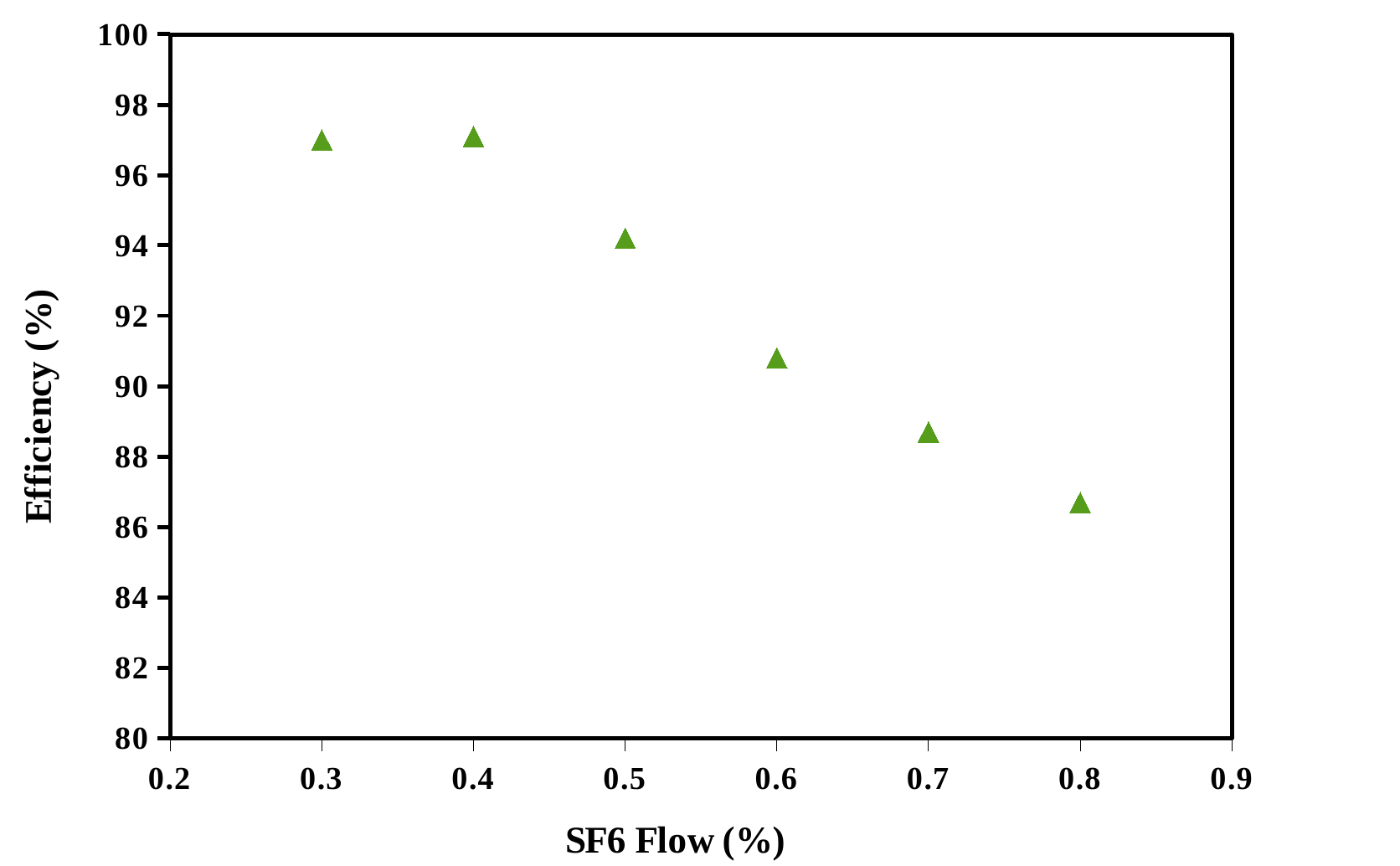}
          \caption{}
          \label{efficiency_vs_SF6}
        \end{subfigure}
        \caption{{\bf (a)} The mean and width of charge distributions. {\bf (b)} The efficiency of RPC. Both were plotted as a function of SF$_6$ concentration.}
    \end{figure}

The charge distributions of the avalanche signals collected at same operating voltage for various proportions of SF$_6$ were fitted with Gaussian's and Landau functions as shown in figure \ref{fig:charge}. In general we have used Landau function for the rising part of the distribution (0-0.5). This Landau function convoluted with the Gaussian which is used for peak as well as tail part of the distribution. Mean and sigma collected from the combined distribution function. As the concentration of SF$_6$ increases from 0.3 $\%$ to 0.8 $\%$, the charge spectra gets narrower and the peak is shifted to the left side. This indicates that only a lesser number of events induces an adequate amount of charge in the strip at higher concentration of SF$_6$.  Thus the mean of the charge distribution decreases with an increase in the SF$_6$ concentration as shown in the figure \ref{mean_charge_vs_SF6}. This behaviour is due to the capture of the electrons in the gas gap by SF$_6$ because of its high electron affinity (1.05 $\pm$ 0.10 eV) \cite{Camarri1998, Salim_SF6}. This reduces the multiplication of electrons in the gas gap. Therefore, the muon events, which cause the production of very less number of primary charge pairs in the gap, would not be able to induce signals above threshold on read-out electrodes on that particular applied high voltage. This results the displacement of working point towards higher field intensity which could increase the noise signals  The efficiency on the working point at 10.4 kV is shown in the figure \ref{efficiency_vs_SF6} which reduced by the addition of SF$_6$ in higher proportions.

    \vspace{-2mm}

\subsection{Time resolution}

\begin{figure}[ht]
        \vspace{-0mm}
        \hspace{-6mm}
        \centering
        \begin{subfigure}[b]{0.33\textwidth}
          \includegraphics[width=1.05\textwidth,trim = 0mm 0mm 0mm 0mm, clip]{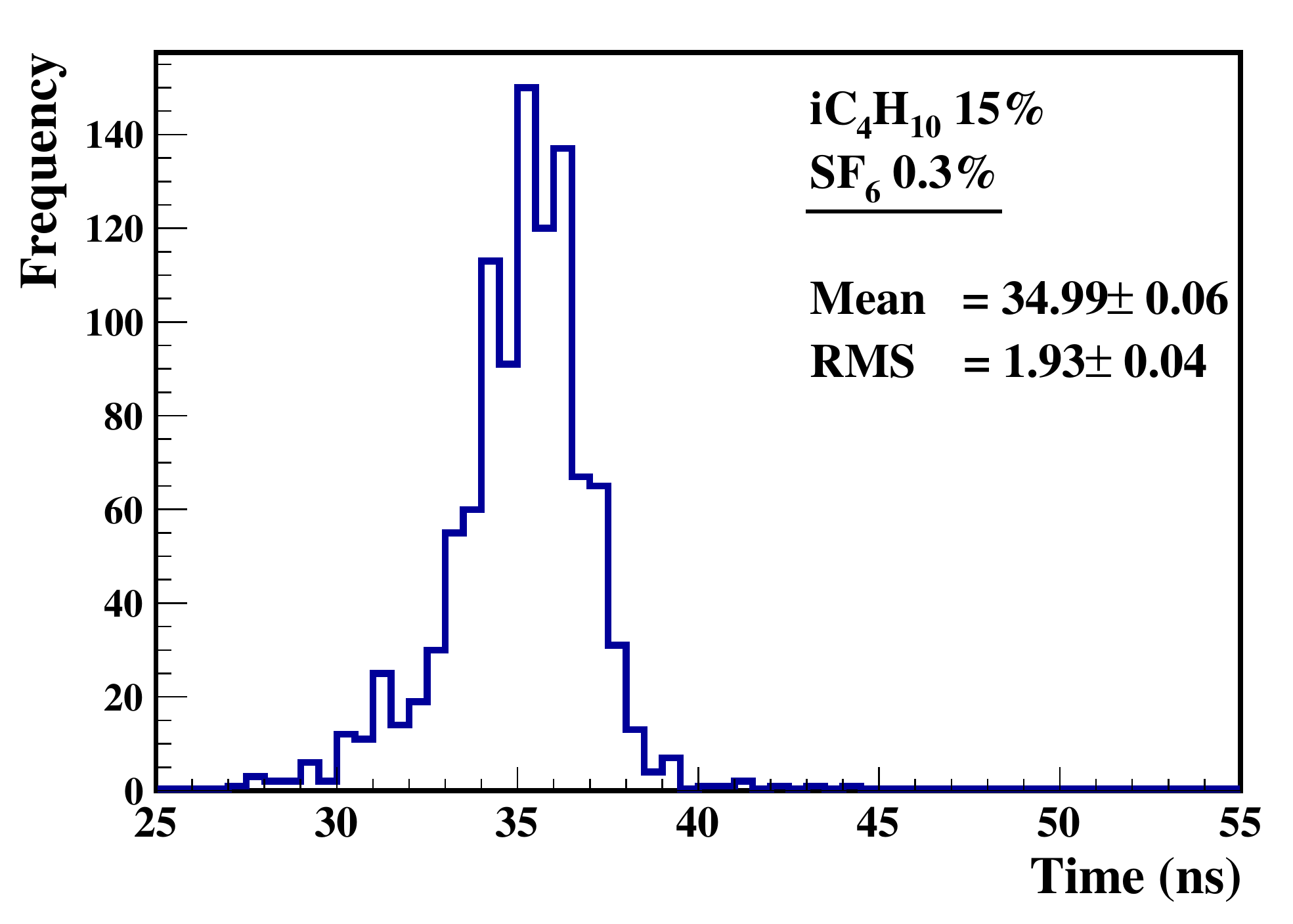}
        \end{subfigure}
        \begin{subfigure}[b]{0.33\textwidth}
          \includegraphics[width=1.05\textwidth,trim = 0mm 0mm 0mm 0mm, clip]{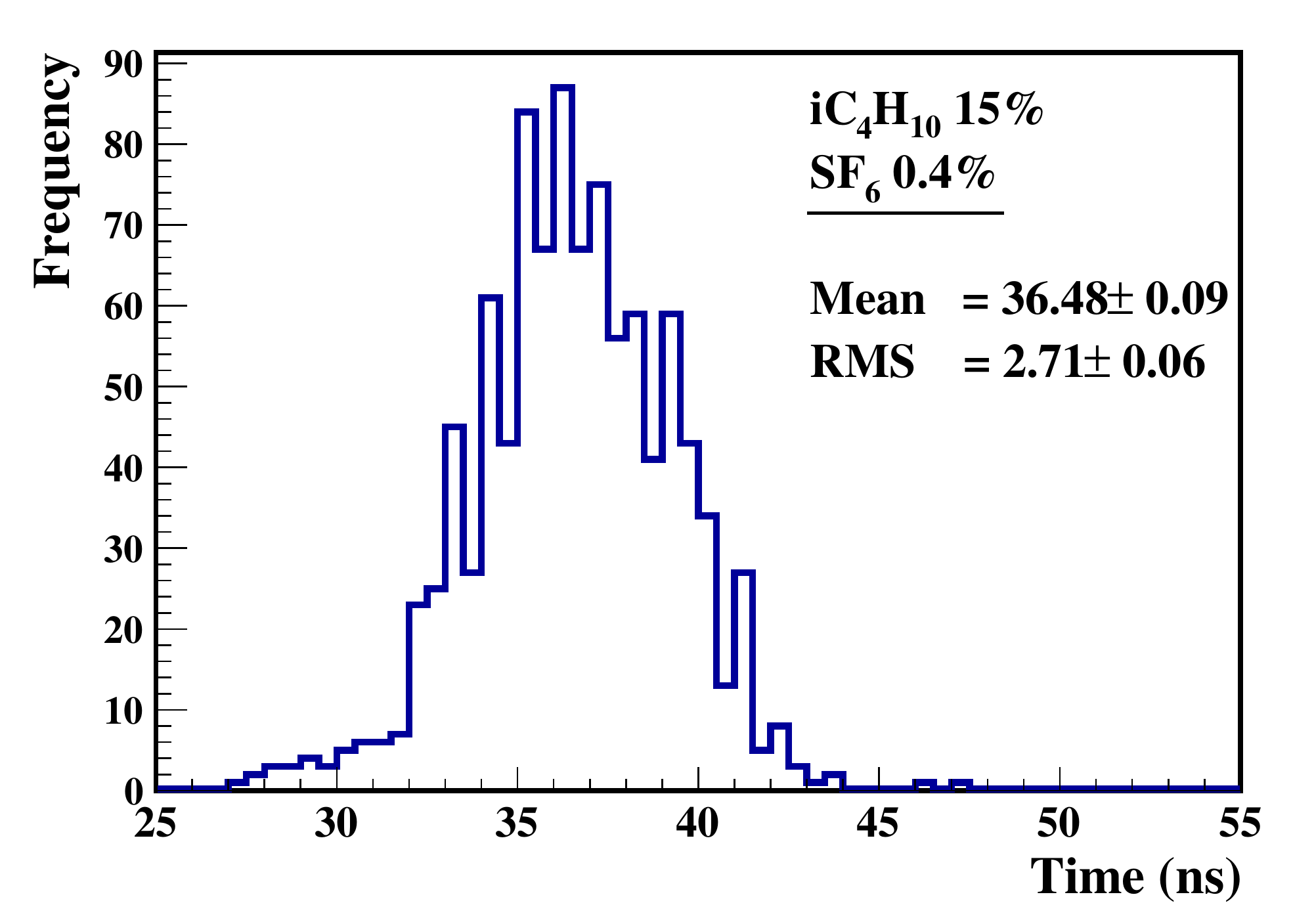}
        \end{subfigure}
        \begin{subfigure}[b]{0.33\textwidth}
          \includegraphics[width=1.05\textwidth,trim = 0mm 0mm 0mm 0mm, clip]{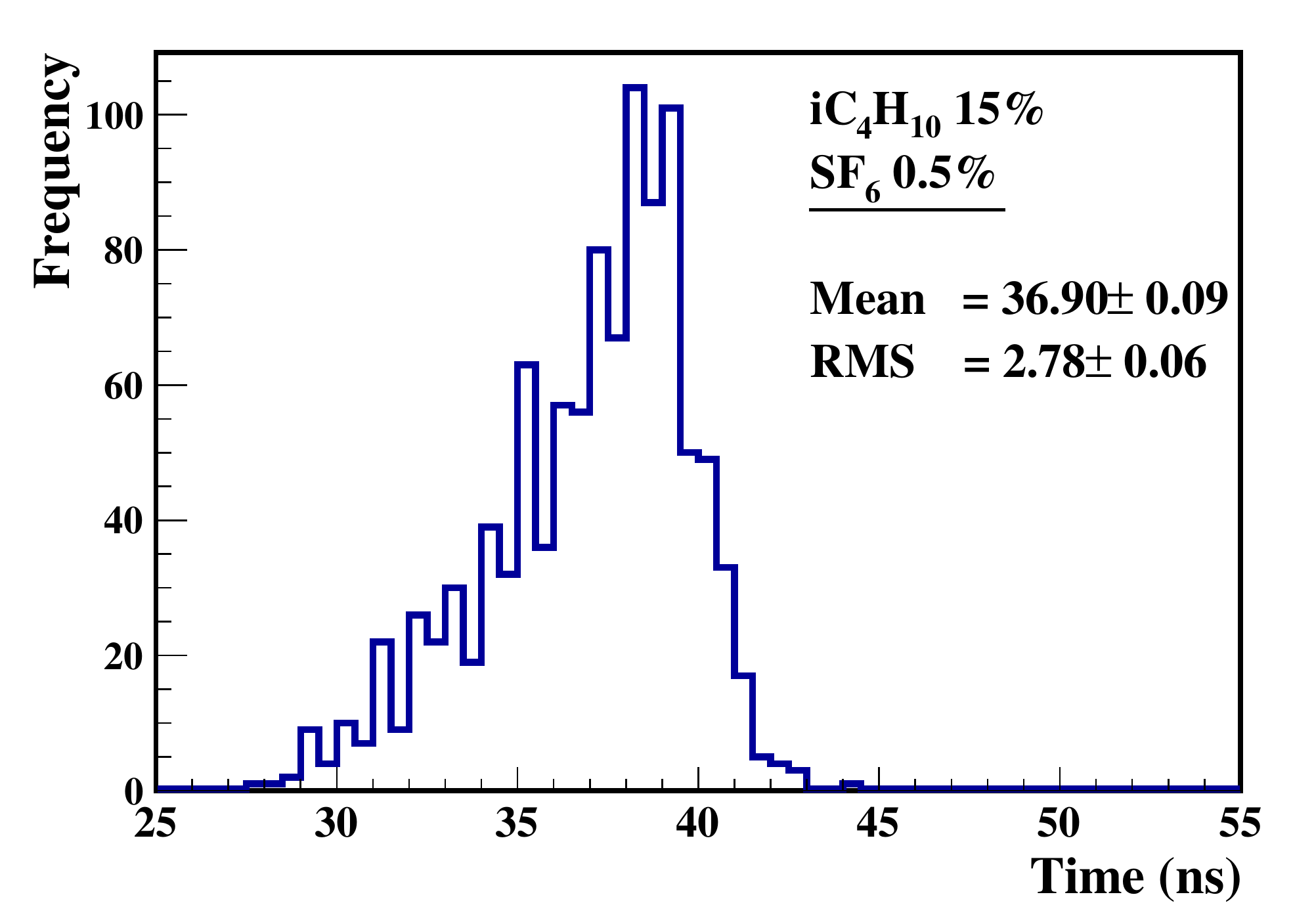}
        \end{subfigure}
        \hspace*{-6mm}
        \begin{subfigure}[b]{0.33\textwidth}
          \includegraphics[width=1.03\textwidth,trim = 0mm 0mm 0mm 0mm, clip]{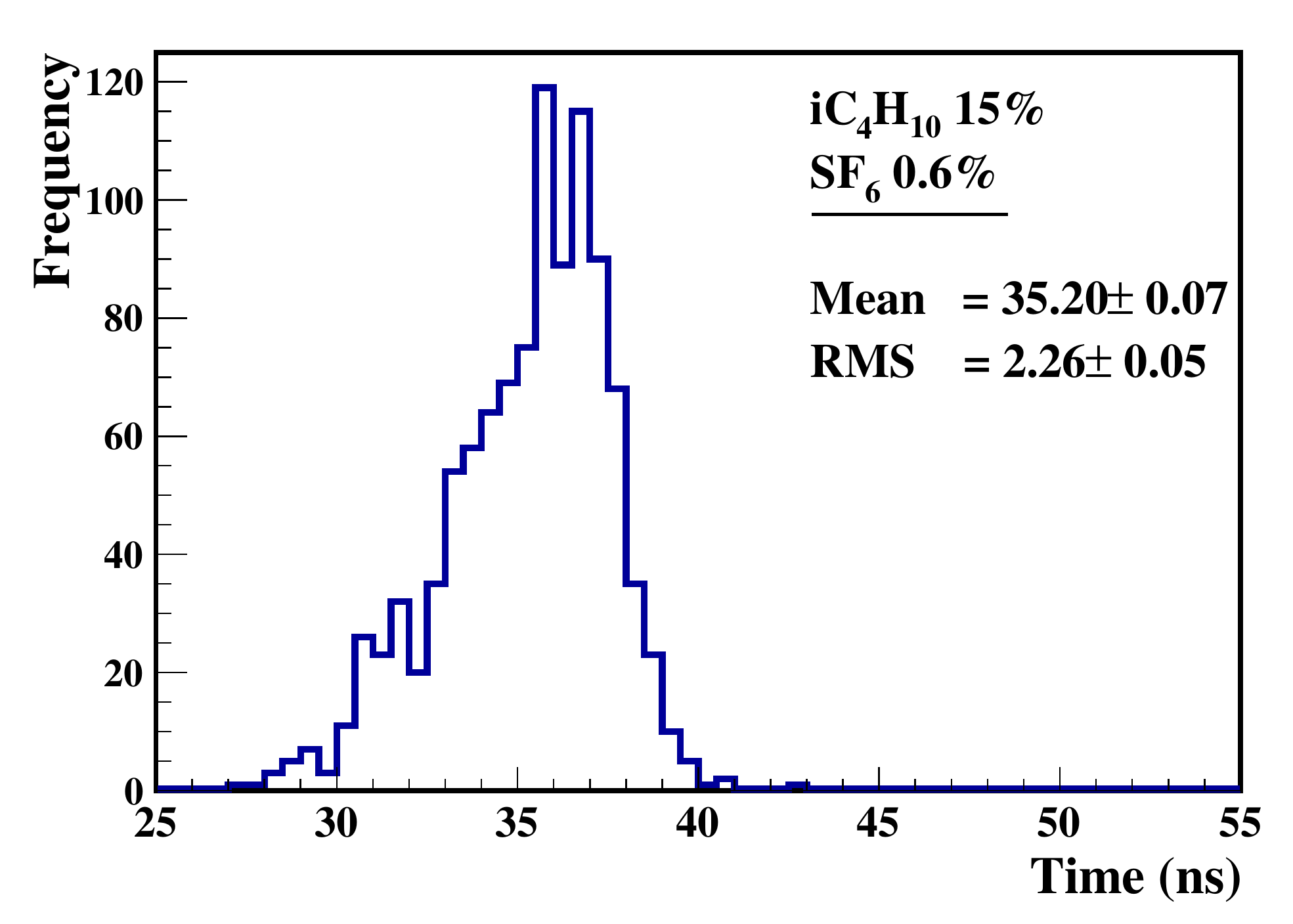}
        \end{subfigure}
        \begin{subfigure}[b]{0.33\textwidth}
          \includegraphics[width=1.03\textwidth,trim = 0mm 0mm 0mm 0mm, clip]{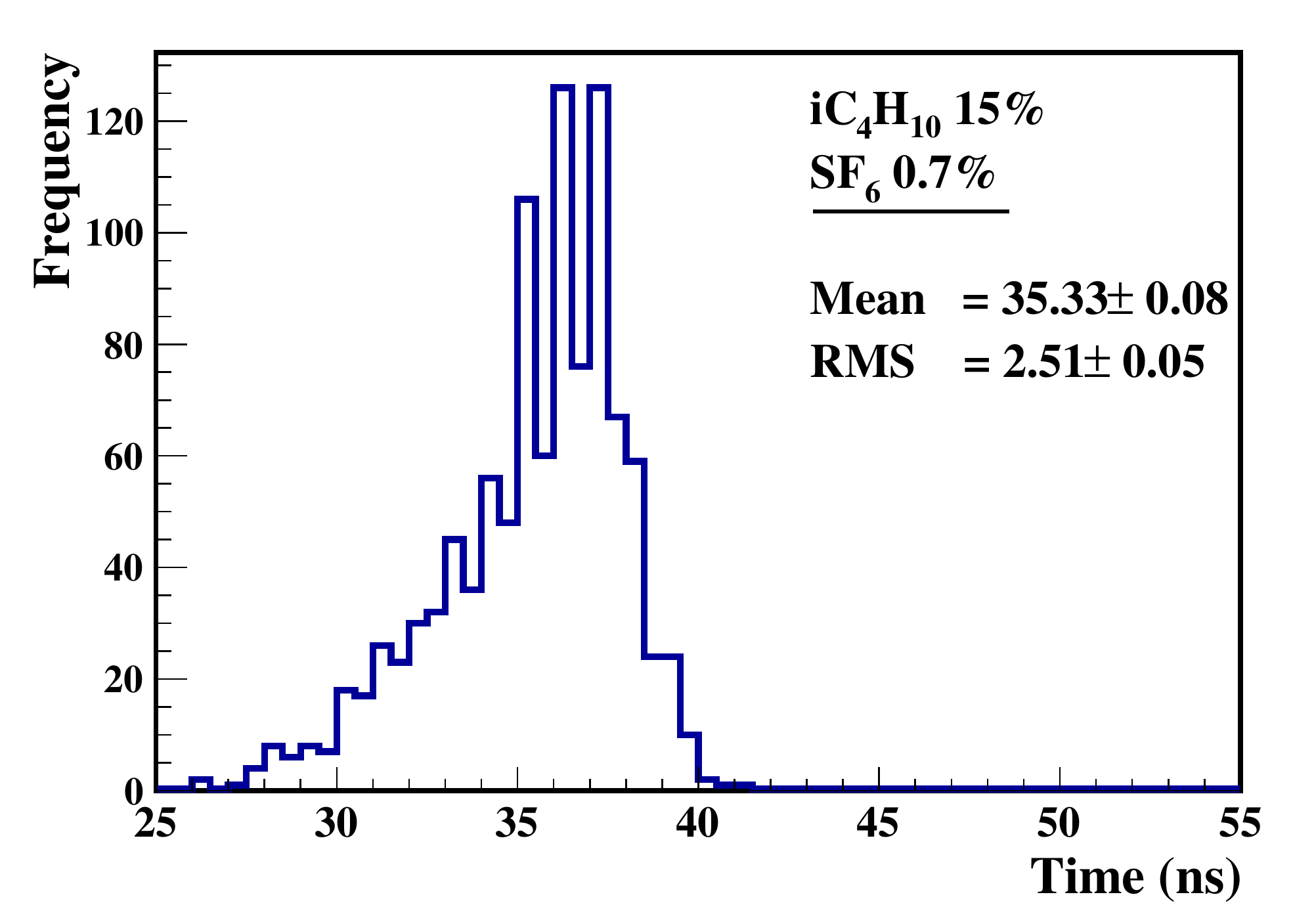}
        \end{subfigure}
        \begin{subfigure}[b]{0.33\textwidth}
          \includegraphics[width=1.03\textwidth,trim = 0mm 0mm 0mm 0mm, clip]{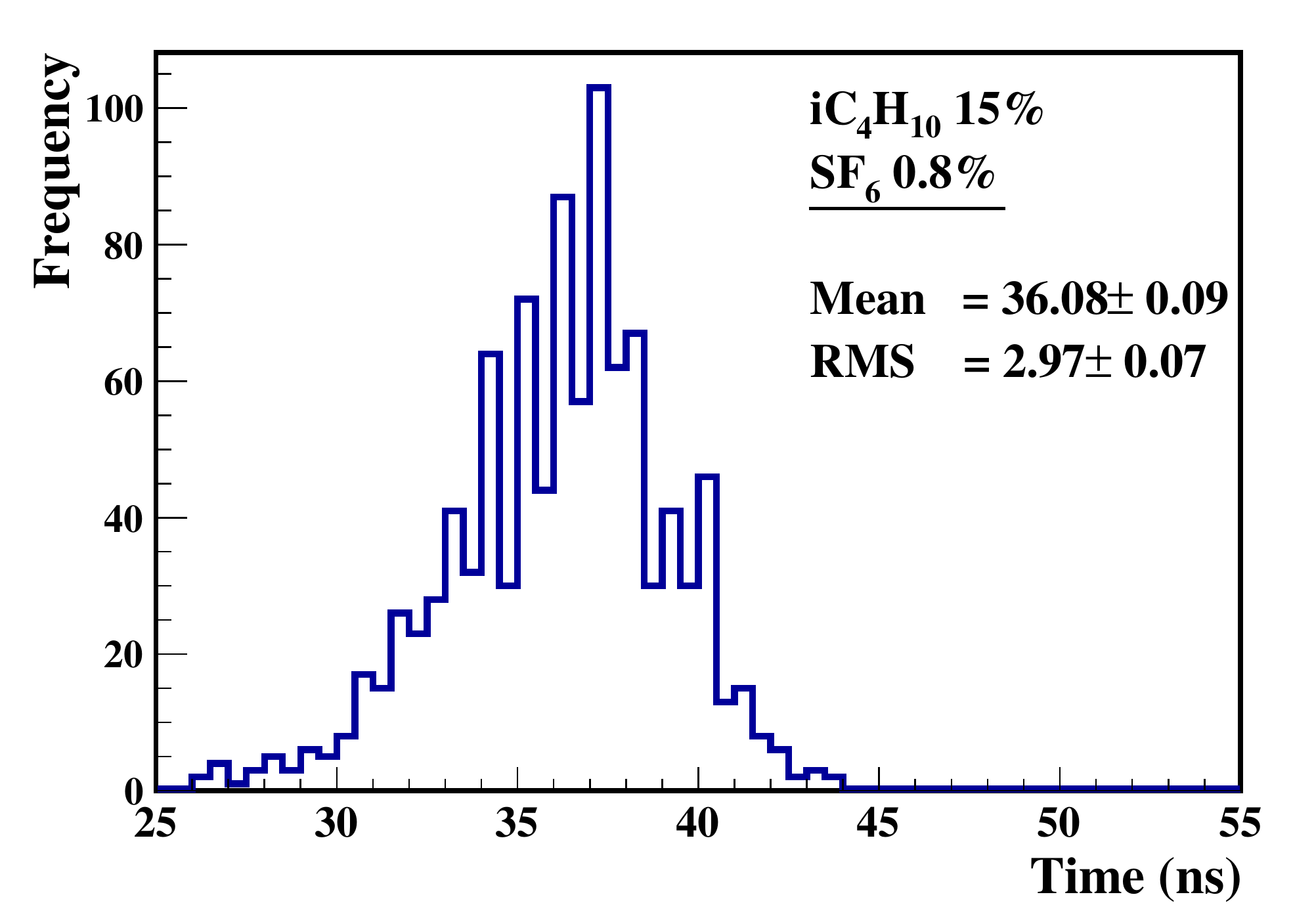}
        \end{subfigure}
        \vspace{-3mm}
        \caption{Time distribution of the RPCs for various SF$_6$ concentrations at same operating voltage.}
        \label{fig:time}
        \vspace{-5mm}
    \end{figure}

The time distribution of the signals collected with RPC at different SF$_6$ proportions is shown in figure \ref{fig:time}. There was no significant effect on the time resolution of the RPCs by the variation in SF$_6$ concentration. 

    \vspace{-3mm}

\newpage
\section{Performance of RPC at different percentages of Isobutane ($\pmb{ iC_4H_{10}}$)}
Isobutane molecules act as quenchers by absorbing UV photons in a wide range of energies and thereby reducing the formation of secondary avalanches. The performance of RPC was studied at various proportions of Isobutane keeping SF$_6$ at 0.4 $\%$. The charge distributions and time resolutions were measured from 1000 signals collected for each proportion of Isobutane. The dependance of average signal charge and time on the Isobutane concentration is analysed. 

\subsection{Charge distribution} 

\begin{figure}[ht]
        \vspace{-2mm}
        \hspace{-2mm}
        \centering
        \begin{subfigure}[b]{0.32\textwidth}
          \includegraphics[width=1.05\textwidth,trim = 0mm 0mm 0mm 0mm, clip]{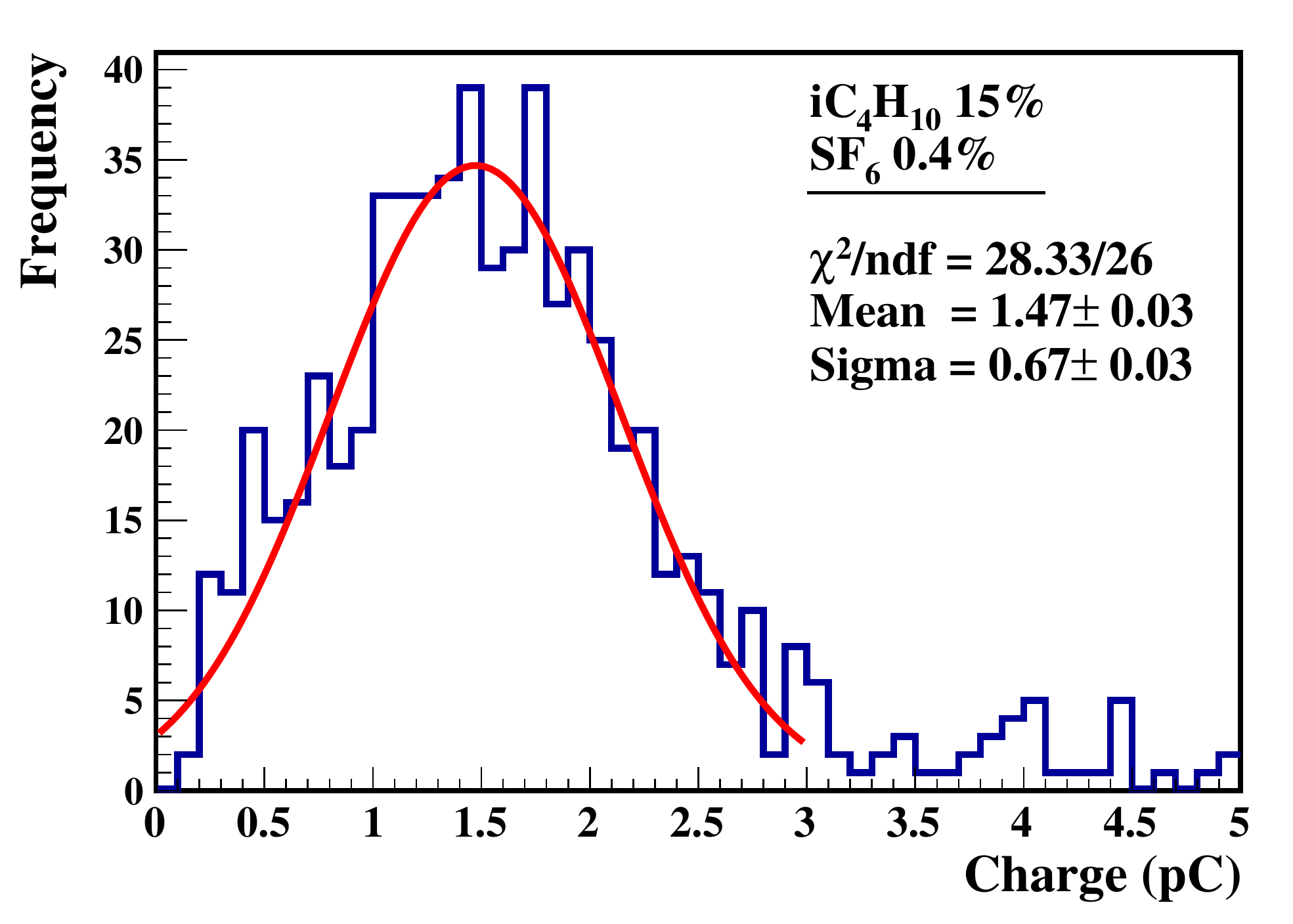}
        \end{subfigure}
        \hspace{0.5mm}
        \begin{subfigure}[b]{0.32\textwidth}
          \includegraphics[width=1.05\textwidth,trim = 0mm 0mm 0mm 0mm, clip]{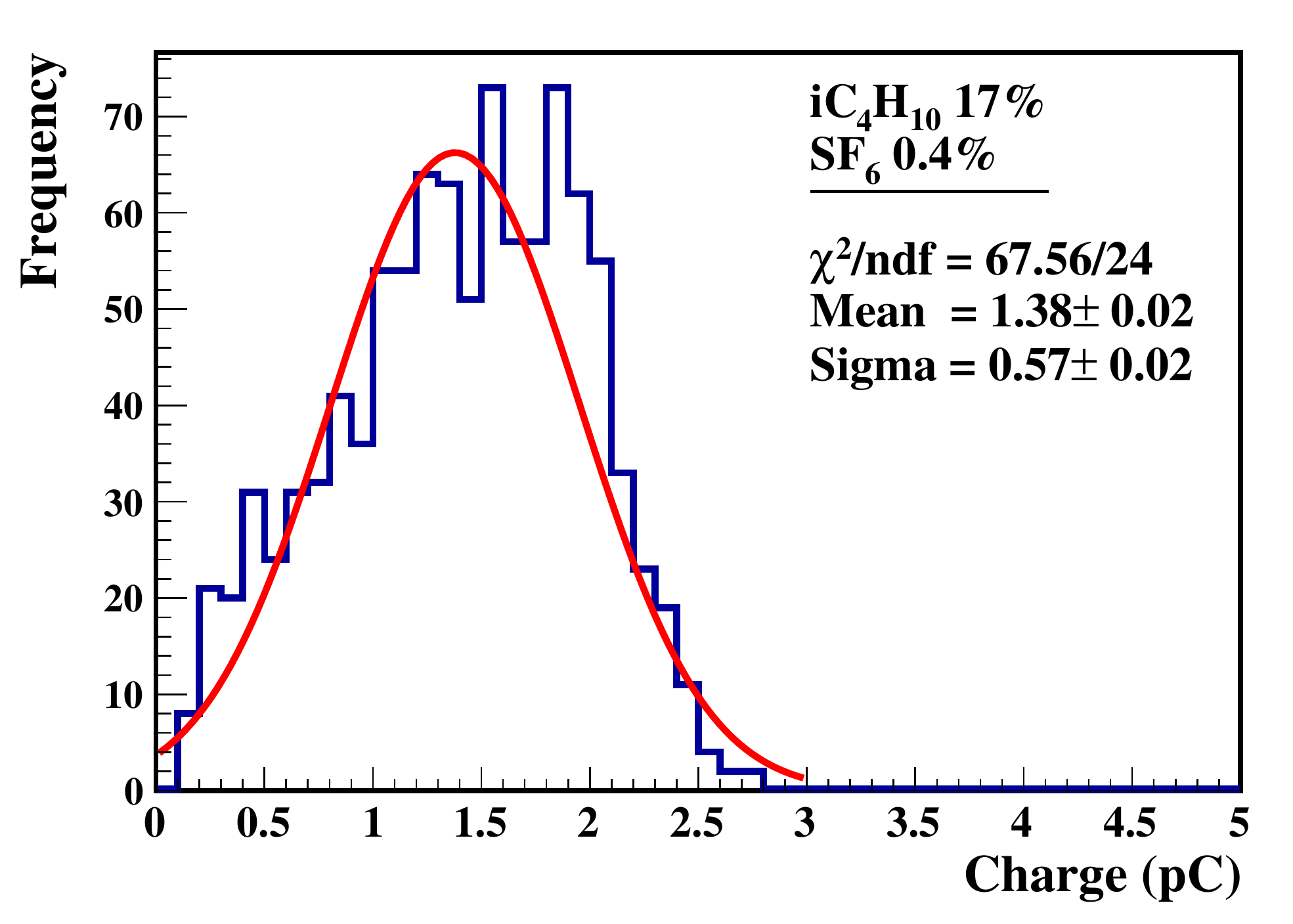}
        \end{subfigure}
        \hspace{0.5mm}
        \begin{subfigure}[b]{0.32\textwidth}
          \includegraphics[width=1.05\textwidth,trim = 0mm 0mm 0mm 0mm, clip]{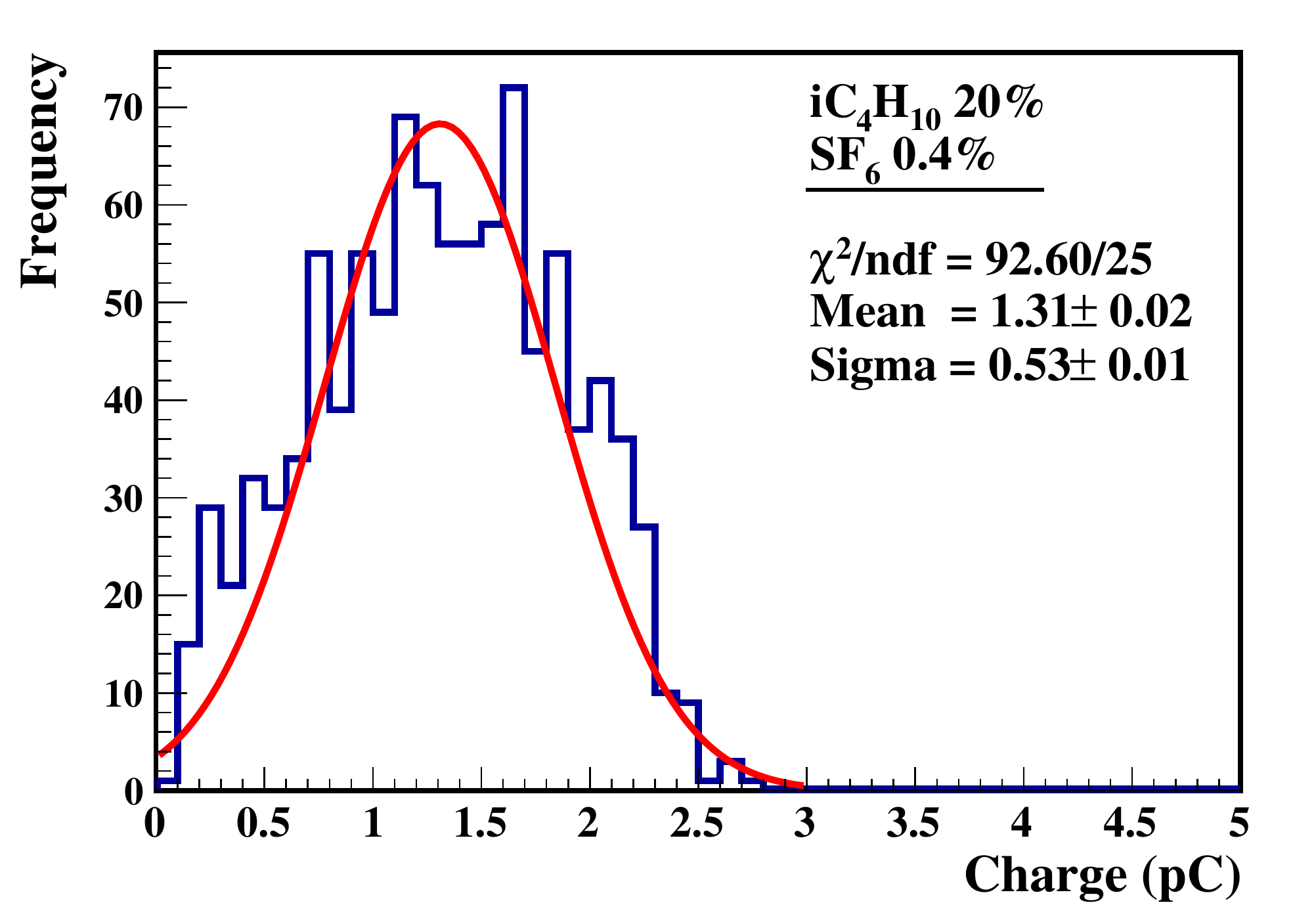}
        \end{subfigure}
        \vspace{-3mm}
        \begin{subfigure}[b]{0.33\textwidth}
          \includegraphics[width=1.05\textwidth,trim = 0mm 0mm 0mm 0mm, clip]{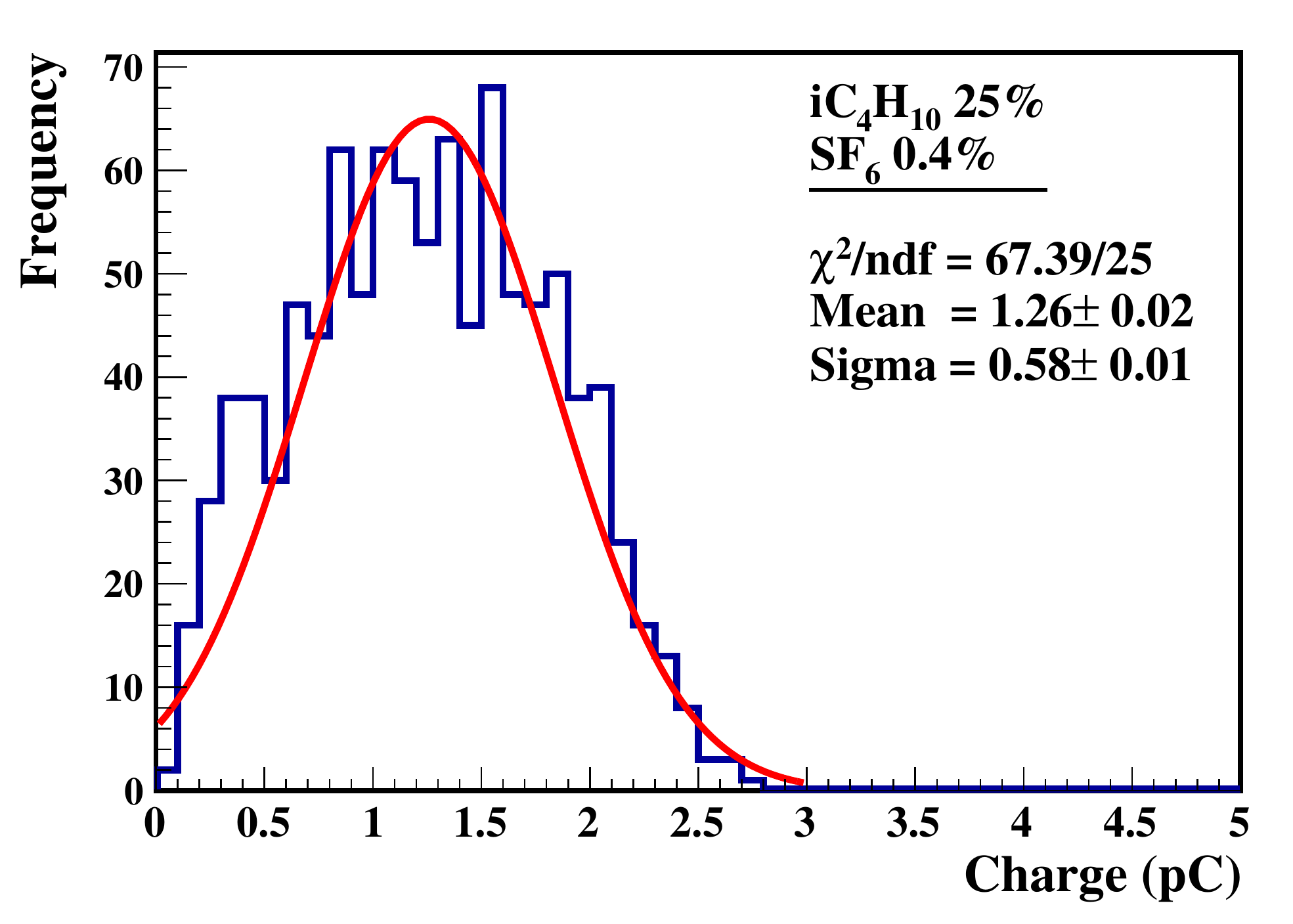}
        \end{subfigure}
        \hspace{0.5mm}
        \begin{subfigure}[b]{0.33\textwidth}
          \includegraphics[width=1.05\textwidth,trim = 0mm 0mm 0mm 0mm, clip]{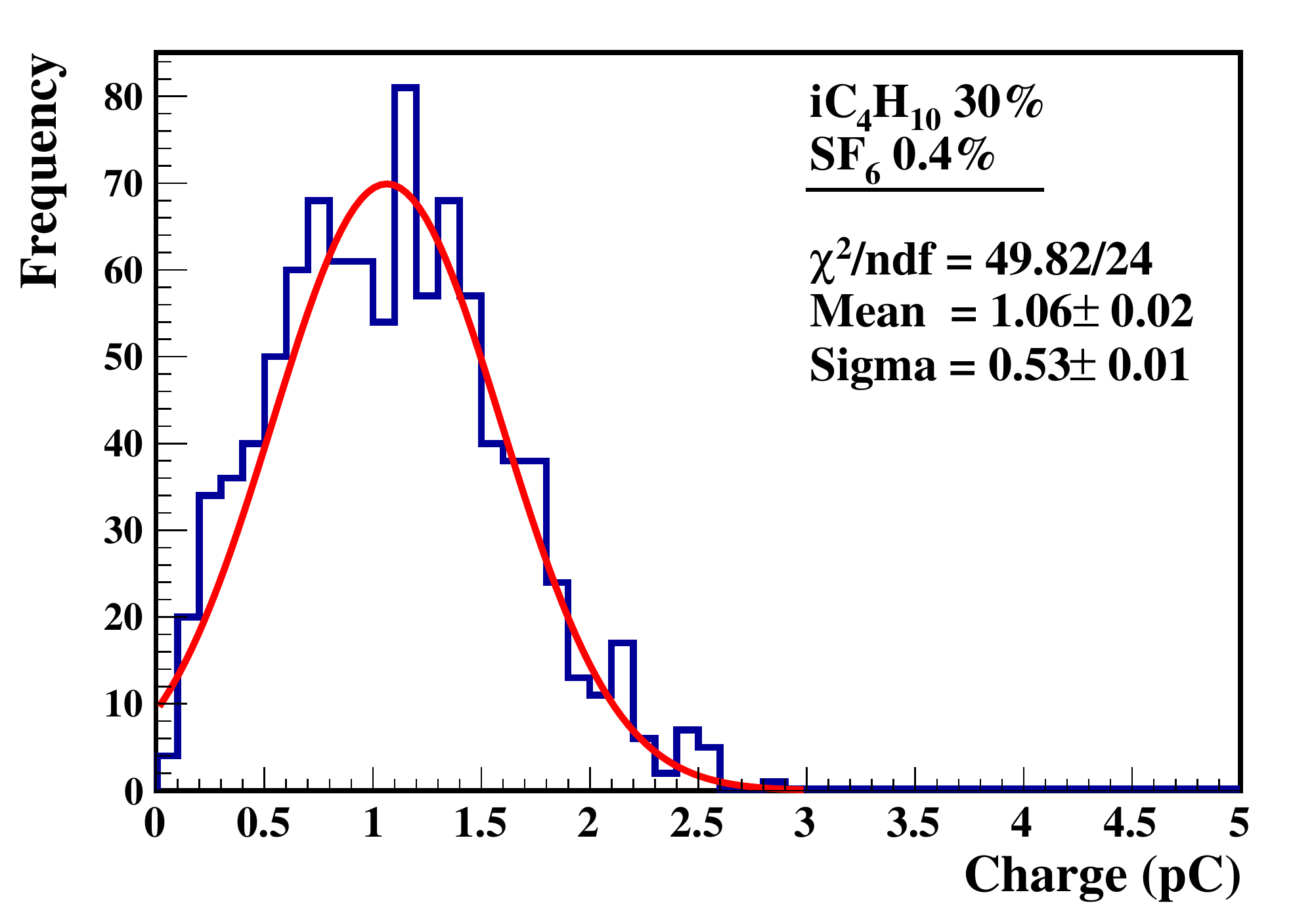}
        \end{subfigure}
        \caption{Charge distribution of the RPCs different Isobutane concentrations at same operating voltage.}
        \label{fig:charge2}
    \end{figure}

\begin{figure}[htbp]
        \hspace{1mm}
        \centering
        \begin{subfigure}[b]{0.48\textwidth}
          \includegraphics[width=1.05\textwidth, height = 43mm, trim = 20mm 163mm 23mm 32mm, clip]{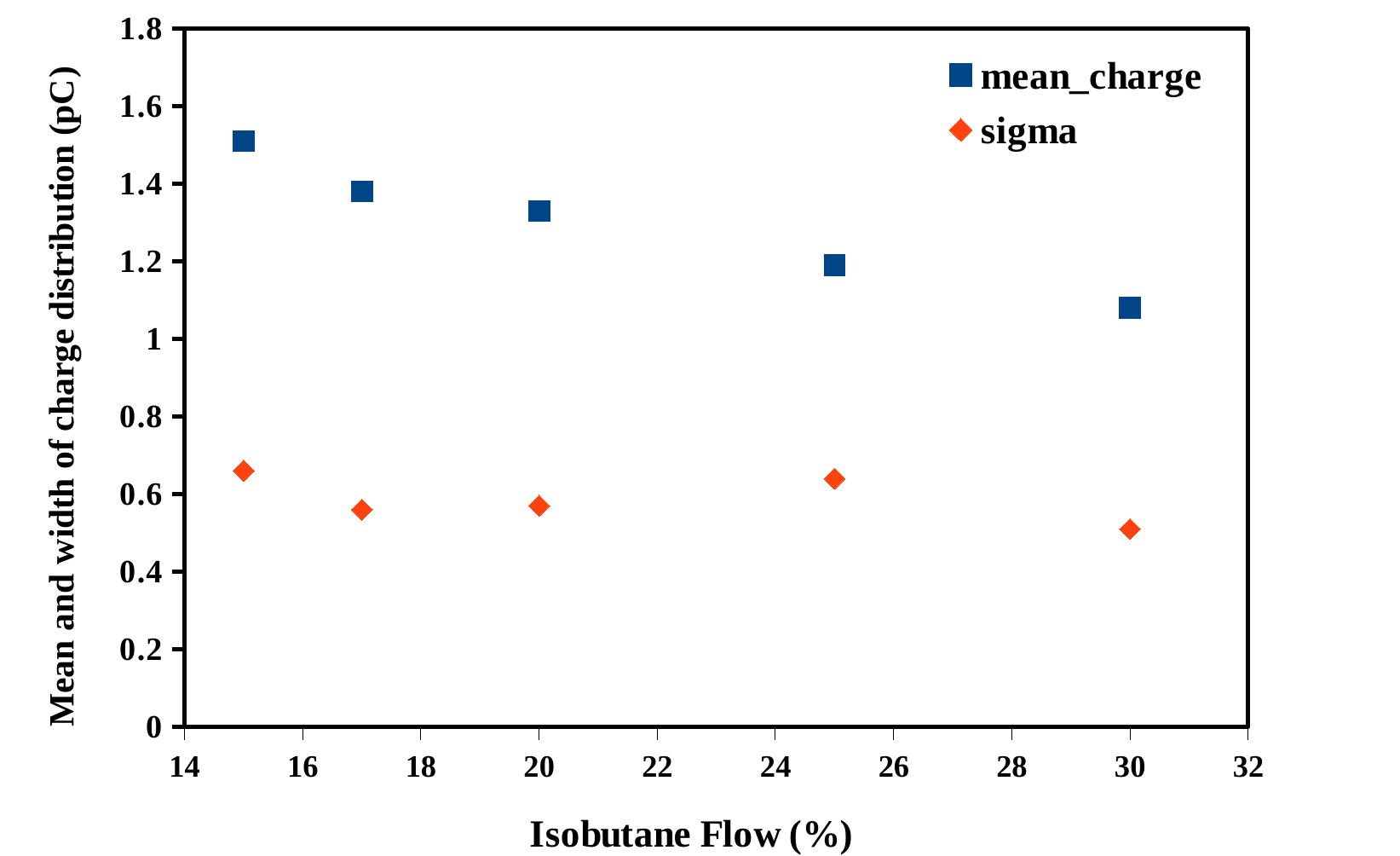}
          \caption{}
        \end{subfigure}
        \hspace{2mm}
        \begin{subfigure}[b]{0.48\textwidth}
          \includegraphics[width=1.05\textwidth, height = 43mm,trim = 20mm 163mm 23mm 32mm, clip]{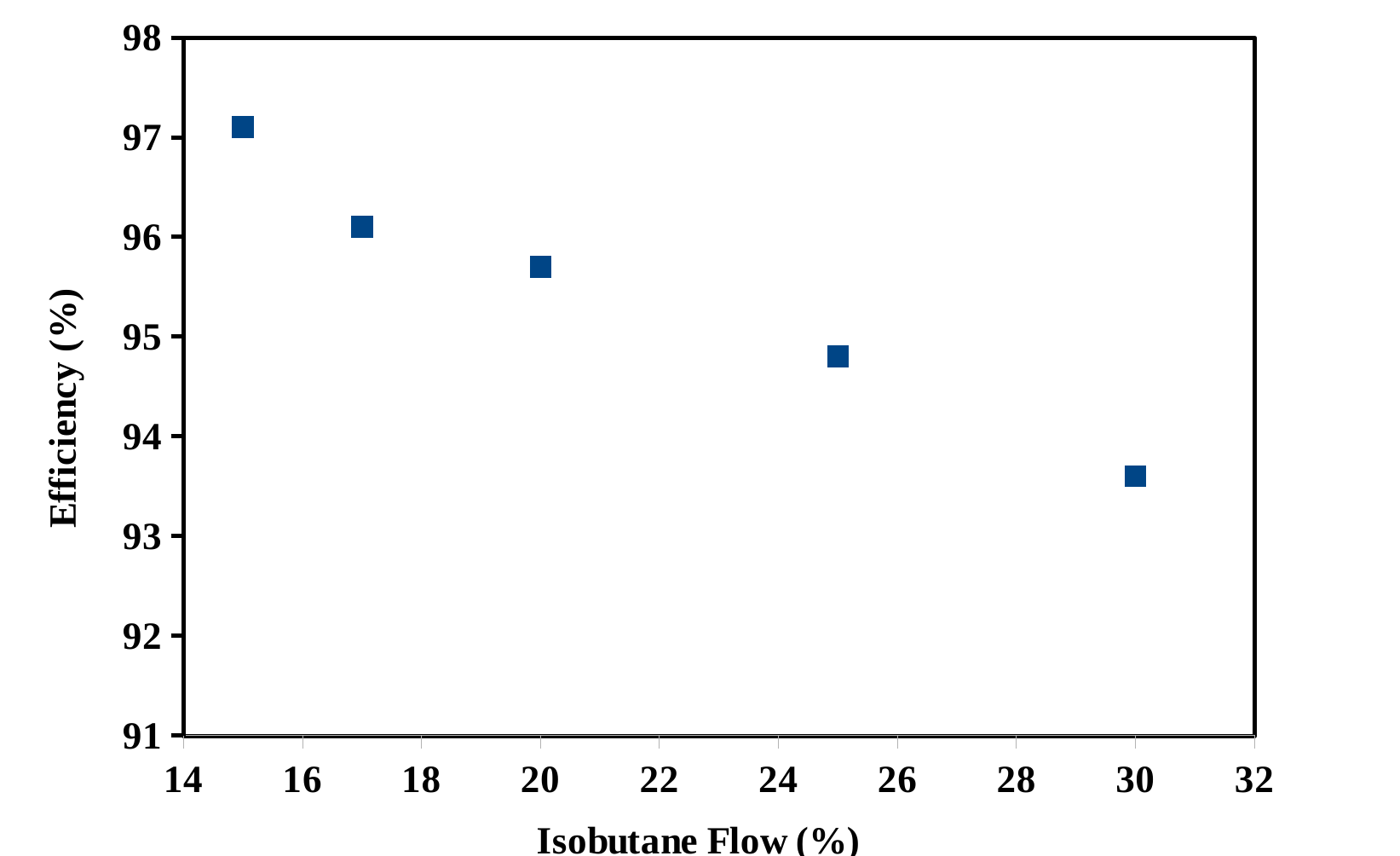}
          \caption{}
        \end{subfigure}
        \vspace*{-9mm}
        \caption{{\bf (a)} The mean and width of charge distributions. {\bf (b)} The efficiency of RPC. Both were plotted as a function of Isobutane concentration.}
        \label{mean_and_efficiency_vs_Isobutane}
    \end{figure}

The charge distributions of the signals collected at various Isobutane proportions were fitted with Gaussians and are shown in figure \ref{fig:charge2}. It is observed from the plot that the RPC is less sensitive to the variations in the proportion of Isobutane in comparison with its sensitivity to SF$_6$ variation. The plot of efficiency versus Isobutane concentration, shown in figure \ref{mean_and_efficiency_vs_Isobutane}, implies that the change in the Isobutane concentration by 10 $\%$ reduces the efficiency by only 2~$\%$ approximately. 

    \vspace{-4mm}

\subsection{Time resolution}

\begin{figure}[ht]
        \vspace{-0mm}
        \hspace{-2mm}
        \centering
        \begin{subfigure}[b]{0.32\textwidth}
          \includegraphics[width=1.05\textwidth,trim = 0mm 0mm 0mm 0mm, clip]{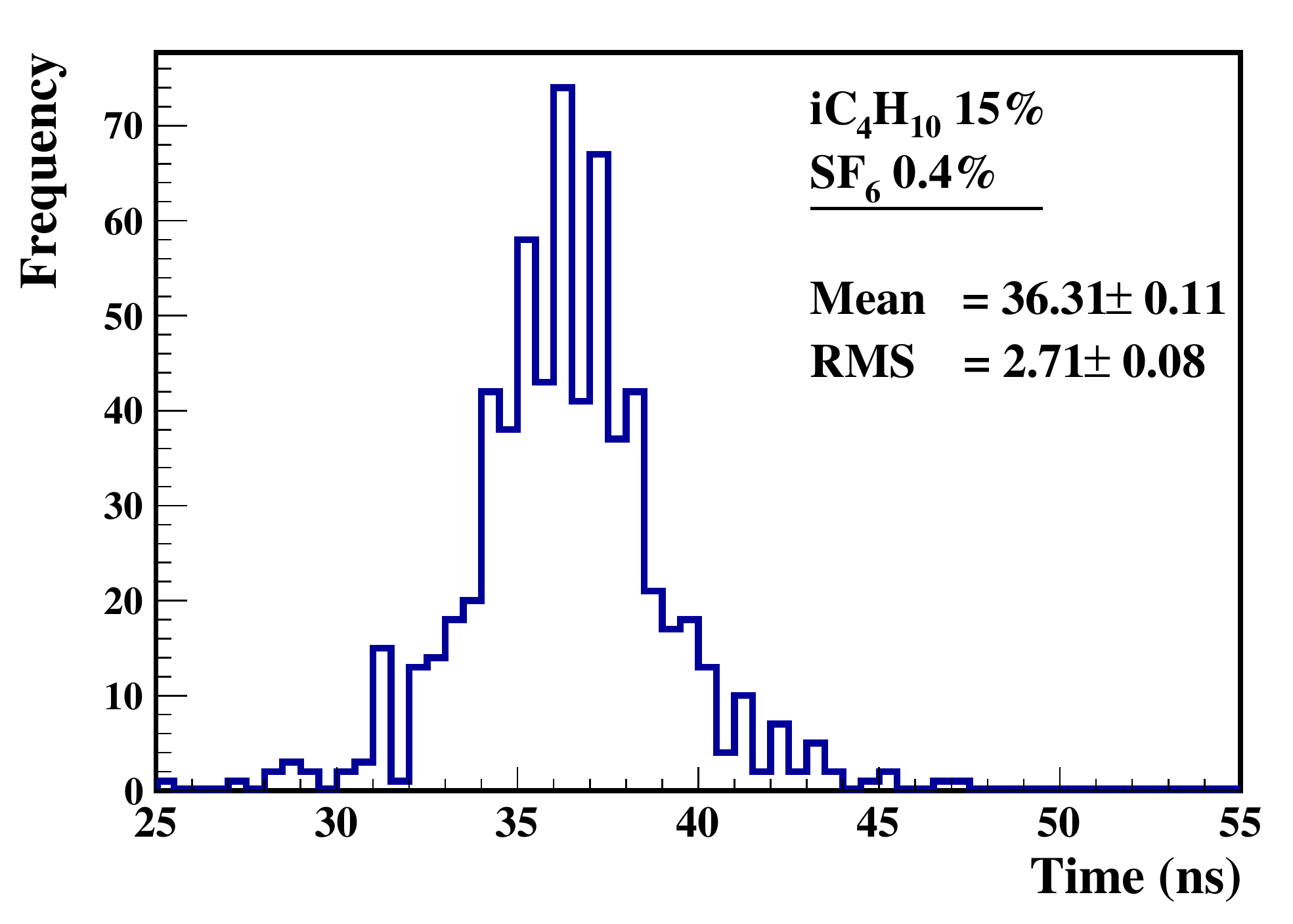}
        \end{subfigure}
        \hspace{0.5mm}
        \begin{subfigure}[b]{0.32\textwidth}
          \includegraphics[width=1.05\textwidth,trim = 0mm 0mm 0mm 0mm, clip]{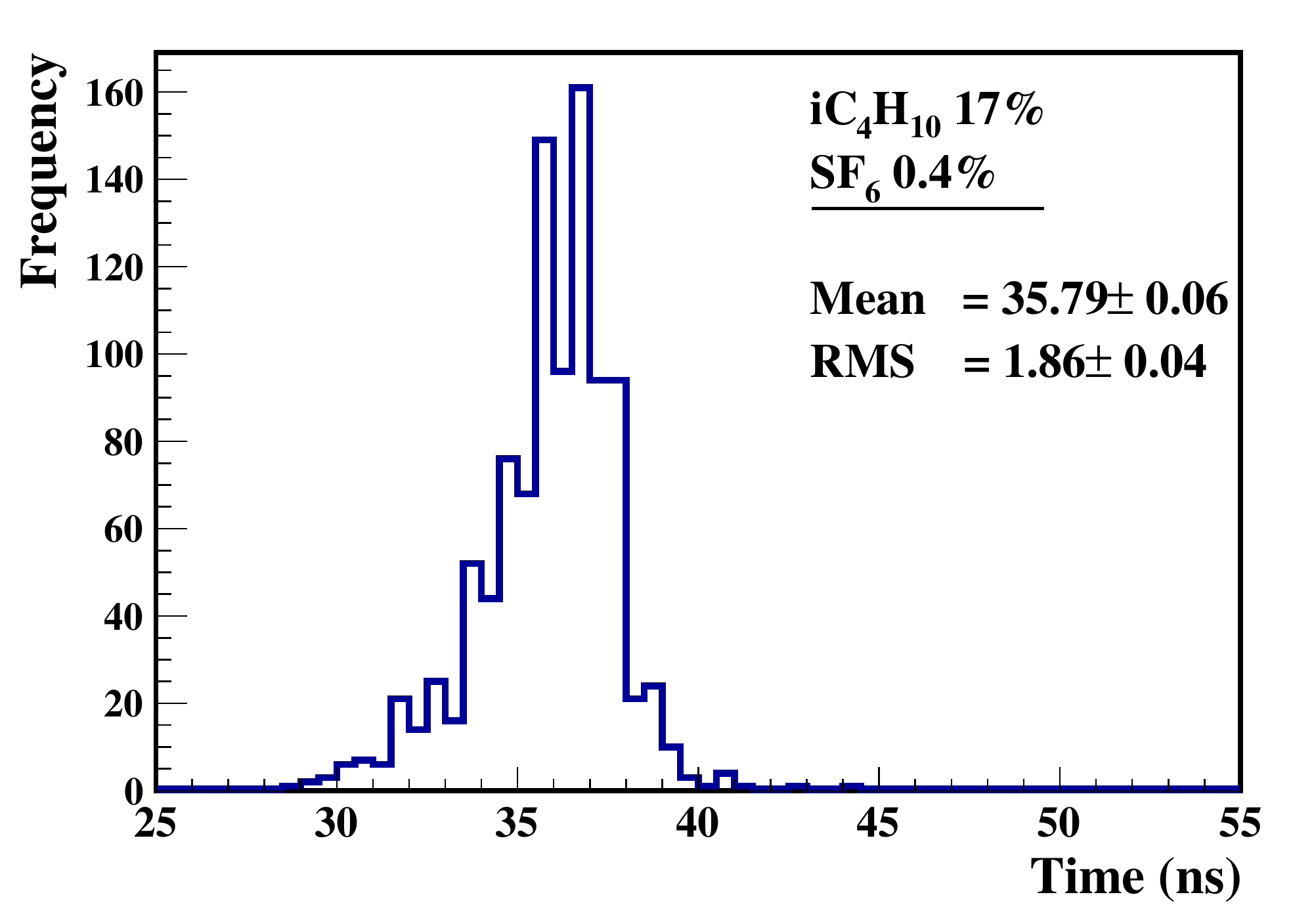}
        \end{subfigure}
        \hspace{0.5mm}
        \begin{subfigure}[b]{0.32\textwidth}
          \includegraphics[width=1.05\textwidth,trim = 0mm 0mm 0mm 0mm, clip]{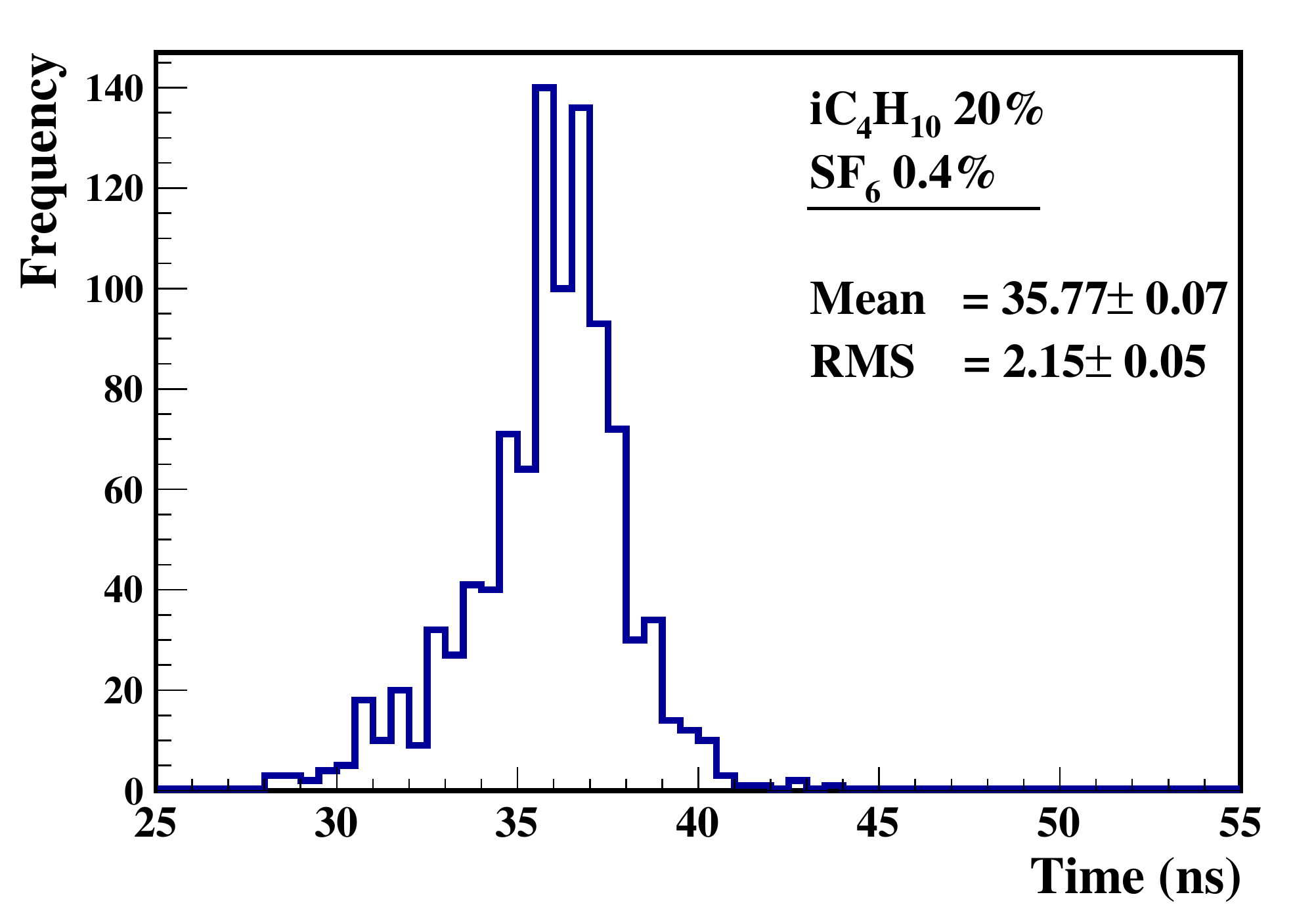}
        \end{subfigure}
        \vspace{-3mm}
        \begin{subfigure}[b]{0.33\textwidth}
          \includegraphics[width=1.05\textwidth,trim = 0mm 0mm 0mm 0mm, clip]{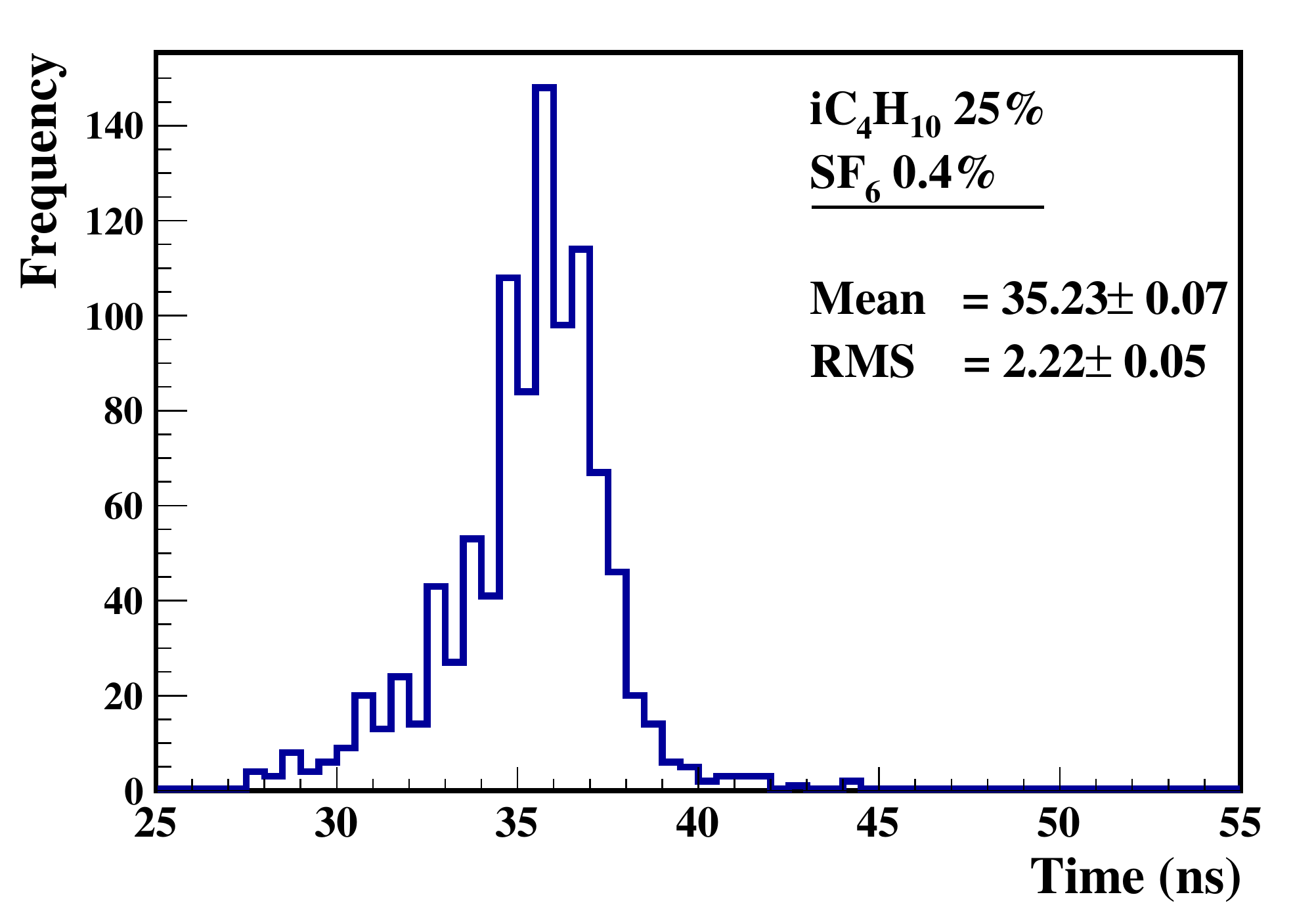}
        \end{subfigure}
        \hspace{0.5mm}
        \begin{subfigure}[b]{0.33\textwidth}
          \includegraphics[width=1.05\textwidth,trim = 0mm 0mm 0mm 0mm, clip]{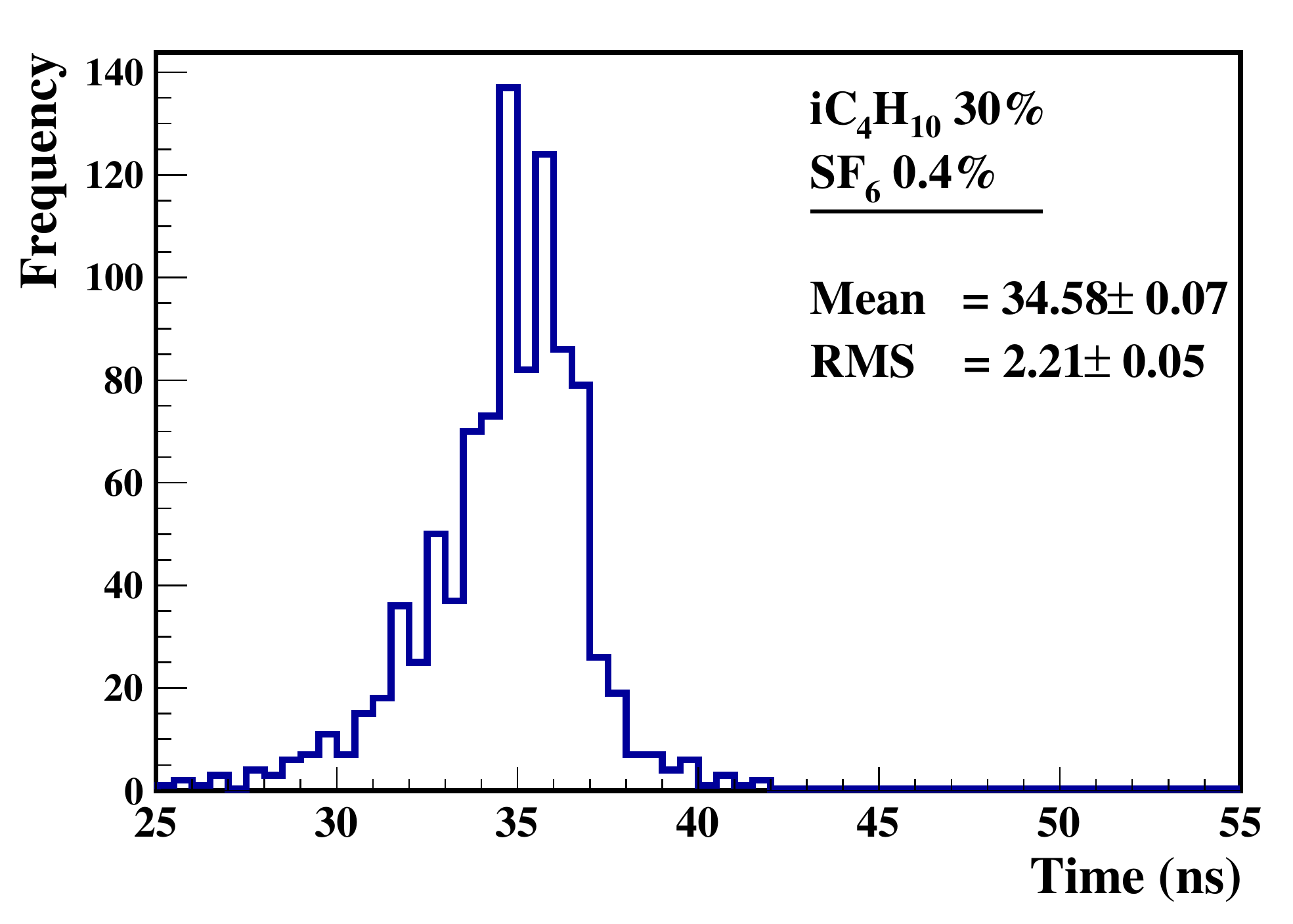}
        \end{subfigure}
        \caption{Charge distribution of the RPCs different Isobutane concentrations at same operating voltage.}
        \label{fig:time2}
        \vspace{-4mm}
    \end{figure}

The time distribution of the signals collected with RPC for various Isobutane proportions is shown in figure \ref{fig:time2}. The time resolution of the RPC is found to be 2 ns approximately. The time resolution is not affected significantly by the variation in Isobutane concentration.

\section{Conclusion}

The variations in the compositions of gaseous ionizing medium in the RPC affects the propagation and multiplication of charge carriers and hence the performance of the RPCs. Therefore, the performance of RPC was studied at different concentration of SF$_6$ and Isobutane.  A slight increase in the SF$_6$ concentration causes significant reduction in the mean of the charge induced on the strips as well as in the efficiency. This is due to the capture of the electrons in the gas gap by SF$_6$ because of its high electron affinity. However, RPC is less sensitive for charge collection to the variations in the concentration of Isobutane compare to the variation of SF$_6$. The mean charge and efficiency were observed to be varied slightly by increasing the concentration of Isobutane gas. This is due to the fact that Isobutane absorbs the UV photons, and hence controls the formation of secondary avalanches. There was no significant difference in the time resolution of RPC by the variation in the concentration of SF$_6$ or Isobutane.

\section*{Acknowledgements}

This work was done with the support of the Department of Atomic Energy (DAE), and the Department of Science and Technology (DST), Government of India. The authors would like to thank INO collaboration for their help throughout the work.


\end{document}